\documentclass[pdftex,epjc3]{svjour3}

\journalname{Eur. Phys. J. C}

\usepackage{graphicx}
\usepackage{amssymb}
\usepackage{amsmath,upgreek,mathtools}
\usepackage[mathscr]{euscript}
\usepackage{widetext}

\newcommand{\lp}{\ell_{\rm p}}
\newcommand{\mpl}{m_{\rm p}}
\newcommand{\gn}{G_{\rm N}}

\def\be{\begin{equation}}
\def\ee{\end{equation}}

\newcommand{\beq}{\begin{equation}}
\newcommand{\eeq}{\end{equation}}

\newcommand{\ba}{\begin{eqnarray}}
\newcommand{\ea}{\end{eqnarray}}

 \def\de{\partial} \def\lb{\rangle}
\def\>{\rangle} 
\def\<{\langle} 
   
\usepackage[colorlinks,hyperindex]{hyperref}
\usepackage{color,float}
\definecolor{green1}{RGB}{0,128,0} 
\hypersetup{
  colorlinks = true,
  linkcolor  = blue,
  citecolor = green1,
}

\usepackage{tensor}
\begin{document}
 
\title{Holographic entanglement entropy under the minimal geometric deformation and extensions}

\author{R. da Rocha\thanksref{e1,addr1}
        \and
        A. A. Tomaz\thanksref{e2,addr1,addr2} 
}

\thankstext[$\star$]{t1}{Thanks to the title}
\thankstext{e1}{e-mail: roldao.rocha@ufabc.edu.br}
\thankstext{e2}{e-mail: anderson.tomaz@ufabc.edu.br}

\institute{Center of Mathematics\label{addr1}\and Center for Natural and Human Sciences\label{addr2}, Federal University of ABC, UFABC, 09210-580, Santo Andr\'e, Brazil
}
\date{Received: date / Accepted: date}

\maketitle

\begin{abstract}
The holographic entanglement entropy (HEE) of the minimal geometric deformation (MGD) procedure, and its extensions (EMGD), is scrutinized within the membrane paradigm of AdS/CFT. The HEE corrections of the Schwarzschild and Reissner--Nordstr\"om solutions, due to a finite fluid brane tension, are then derived and discussed in the context of the MGD and the EMGD. 
\end{abstract}


\section{Introduction}
\label{intro}
\quad\; The AdS/CFT duality generally states that weakly-coupled gravity in $(d + 1)$-dimensional anti-de Sitter (AdS) space is the theory dual to a strongly-coupled conformal field theory (CFT), whose underlying hydrodynamical limit corresponds to the Navier--Stokes equations -- at the $d$-dimensional AdS boundary \cite{Maldacena:1997re,Bilic:2015uol,hub}. The membrane paradigm is usually deployed into the fluid/gravity correspondence,  as a low-energy regime of AdS/CFT  \cite{Eling:2009sj}. 
 In the membrane paradigm setup, black holes were studied  in the infrared (IR) limit \cite{Casadio:2015gea,daRocha:2017cxu,Ovalle:2017wqi}.  In addition, the seminal Refs. \cite{maartens,Antoniadis:1998ig,Antoniadis:1990ew} present 
 important features of this duality. For a $N$ number of colours, indexing a SU($N$) (gauge) theory, AdS/CFT duality asserts that $\mathcal{N} = 4$ superconformal Yang--Mills theory in 4D is dual to type IIB string theory on AdS$_5 \times S^5$. In the original setup, the AdS$_5$ boundary is a 4D  Minkowski spacetime, and the $D_3$-brane near horizon geometry is the AdS$_5$ space, whereas the  far away brane geometry remains flat. 
  
 In the membrane paradigm of AdS/CFT, encompassing General Relativity (GR),  the so called method of geometric deformation (MGD)  
places itself as an important procedure to generate new solutions of the effective Einstein's field equations on the brane \cite{Casadio:2015gea,Ovalle:2017wqi,Ovalle:2017fgl,covalle2,Ovalle:2014uwa,Ovalle:2016pwp}, including anisotropic solutions, 
describing  compact stellar distributions, in a Weyl fluid flow in the bulk  \cite{Ovalle:2007bn,Casadio:2015jva}. 
The MGD and its extensions take into account the brane Einstein's field equations \cite{GCGR,CoimbraAraujo:2005es}, where the effective stress-energy tensor has additional terms, in particular regarding the Gauss--Codazzi equations from the bulk stress-energy tensor projected onto the brane \cite{maartens}. Important terms, constituting the effective brane stress-tensor, are the bulk dark radiation, the bulk dark pressure, the electric part of the Weyl tensor and quadratic terms on the brane stress-energy tensor. This last one is derived for regimes of energy that are beyond the (finite) brane tension in the theory. Being  our universe described by a brane with tension
$\sigma$, the MGD leads to a deformation of the  Schwarzschild metric proportional to a positive length
scale $\ell\sim\sigma^{-1}$ \cite{Ovalle:2017wqi,Ovalle:2017fgl,covalle2}.

 The MGD and its extensions \cite{Casadio:2015gea} have been recently equipped with experimental, phenomenological, and  observational very precise bounds, physically constraining their running parameters. MGD gravitational lensing effects were explored in Ref. \cite{Cavalcanti:2016mbe} and the classical tests of GR imposed bounds on the brane tension in Ref. \cite{Casadio:2015jva}. The most precise values of the brane tension range were obtained in Refs. \cite{Casadio:2016aum,Fernandes-Silva:2019fez}. In fact, in these references, the information entropy was used to provide account for the critical stellar densities, in the MGD and EMGD setups, deriving analogue of the Chandrasekhar's critical stellar densities,  that are also extremal points of the system associated configurational entropy \cite{Casadio:2016aum,Fernandes-Silva:2019fez}. Besides, MGD black hole  analogues were explored in Ref. \cite{daRocha:2017lqj}. Sound waves into and out of  de Laval nozzles derives experimental data about the bulk Weyl fluid. Acoustic perturbations in MGD nozzles were shown to play the role of MGD quasinormal modes. Besides, MGD black branes was also studied in Ref. \cite{Casadio:2013uma} and 2+1 MGD solutions were scrutinized in Ref. \cite{Contreras:2018vph}. Ref. \cite{Contreras:2018gzd} showed that any static and spherically symmetric anisotropic solution of the
Einstein's field equations can be thought of as being a system sourced by certain deformed isotropic system, 
in the context of MGD approach. 
Anisotropic MGD solutions were obtained in Refs. \cite{PerezGraterol:2018eut,Heras:2018cpz,Morales:2018urp} and \cite{Fernandes-Silva:2017nec}. Besides, anisotropic MGD-like solutions were obtained by gravitational decoupling \cite{Ovalle:2017wqi,Ovalle:2017fgl,Gabbanelli:2018bhs,Panotopoulos:2018law}, whereas   conformal sectors were analyzed in Ref. \cite{Ovalle:2017khx}. 
The MGD was also used to study bulk effects on realistic stellar interior distributions~\cite{Ovalle:2013xla} and the in the analysis of hydrodynamics of
black strings, in the AdS/CFT membrane paradigm~\cite{Casadio:2013uma}. 
Recently, the MGD corrections to the gravitational lensing was estimated in Ref.~\cite{Cavalcanti:2016mbe},
and it was shown that the merging of MGD stars may be easier  detected by the eLISA experiments, when compared with their Schwarzschild counterparts~\cite{daRocha:2017cxu}. 
MGD black strings were shown to be stable under small linear perturbations~\cite{Fernandes-Silva:2017nec}. EMGD stellar distributions were also employed to study dark hidden gauge sectors, 
in the context of glueballs stars, and their observational signatures in Ref. \cite{Fernandes-Silva:2018abr}. Besides, the MGD was employed in the context of the generalized uncertainty principle, where Hawking fermions were analyzed \cite{Casadio:2017sze}.  

Another relevant setup, primarily motivated to describe black hole physics, is  
entanglement entropy (EE), that has been explored in  several fields. Here the AdS/CFT correspondence setup will be employed 
in this context. One can investigate how to approach the inverse problem to that one solved in Ref. \cite{Ryu:2006bv}, namely how to use the entanglement entropy for a given quantum system to reconstruct the geometry of the corresponding bulk.
The holographic entanglement entropy (HEE) was employed to compute the entanglement entropy of a subsystem in the dual theory. When the bulk theory is the Einstein's gravity, the HEE was conjectured, for a subsystem on the boundary, to be identical to the Bekenstein--Hawking formula, relating the area  of a minimal surface that has the entangling surface as its own boundary. As the so-called Ryu and Takayanagi formula involves a minimal surface, it is important to analyze such minimal surfaces in various asymptotically AdS spacetimes \cite{Hubeny:2007xt,Emparan:2006ni}. The 
HEE derivation can be found in Ref. \cite{Casini:2011kv}. Our main aim in this paper is to emulate previous formulations of the HEE and apply the MGD and the EMGD in this context, therefore scrutinizing 
the physical consequences and their deviations from the Schwarzschild and Reissner--N\"ordstrom (RN) solutions as well. 

This paper is organized as follows: in Sect.~\ref{sect:MGD-setup} we promote a general review of the MGD and EMGD setup. The HEE for spherically symmetric spacetimes anchored in the Ryu-Takayanagi formula is then briefly presented. The computations of the HEE corrections for a MGD spacetime is described and showed in Sect.~\ref{sect:MGD} either with boundaries far from the event horizon or almost on it. In Sect.~\ref{sect:EMGD} we develop the computation of the HEE corrections for EMGD spacetimes. Further discussions, analysis, conclusions and perspectives are outlined in Sect.~\ref{sect:Finals}.

\section{The MGD setup in the membrane paradigm}
\label{sect:MGD-setup}
The MGD procedure can be realized as a mechanism that is usually employed to derive high energy corrections to the GR. The MGD is a well-established method that controls the strong non-linearity of Einstein's  field equations, with more intricate stress-energy tensor, in such a way not to produce inconsistencies in the obtained gravitational solutions. The MGD is naturally seen into the AdS/CFT correspondence, which can bind higher-dimensional models to 4D theories that are strongly-coupled. According to the membrane paradigm of AdS/CFT, that has been used to realize the deformation method, a finite brane tension plays the role of the brane energy density, $\sigma$. There is a fine-tuning between $\sigma$, and the running brane and bulk cosmological parameters \cite{maartens}. 
Systems with energy $E\ll\sigma$ neither feel the self-gravity effects nor the bulk effects, which  then  allows the recovery of GR in such a regime. An infinitely rigid brane scenario, representing the 4D brane manifold, can be  implemented in the  $\sigma\to\infty$ limit. 
The most strict brane tension bound, $\sigma \gtrsim  2.83\times10^6 \;{\rm MeV^4}$, was derived in the extended MGD (EMGD) context in Ref. \cite{Fernandes-Silva:2019fez}. 

The Gauss--Codazzi equations can be used to represent the brane Ricci tensor to the bulk 
geometry, when the discontinuity of the extrinsic curvature is related to the brane stress-tensor. Hence, the bulk field equations \cite{GCGR} yield the effective Einstein's field equations on the brane, whose corrections consist of a byproduct of an AdS bulk Weyl fluid. This fluid flow is implemented by the bulk Weyl tensor, whose projection onto the brane, the so-called electric part of the Weyl tensor, reads    
\begin{eqnarray}
\!\!\!\!\!\!\!\!\mathcal{E}_{\mu\nu}(\sigma^{-1}) \!=\!-6\sigma^{-1}\!\left[ \mathcal{U}\!\left(\!u_\mu u_\nu \!+\! \frac{1}{3}h_{\mu\nu}\!\right) \!+\! \mathit{Q}_{(\mu} u_{\nu)}\!+\!\mathcal{P}_{\mu\nu}\right], \label{A4}
\end{eqnarray}
\noindent where $h_{\mu\nu}$ denotes the projector operator onto the brane that is orthogonal to the  $4$-velocity, $u^\mu$, associated to the  Weyl fluid flow. Besides, $\mathcal{U}=-\frac16\sigma\mathcal{E}_{\mu\nu} u^\mu u^\nu$ is the effective energy density; $\mathcal{P}_{\mu\nu}=-\frac16\sigma\left(h_{(\mu}^{\;\uprho}h_{\nu)}^{\;\sigma}-\frac13 h^{\uprho\sigma}h_{\mu\nu}\right)\mathcal{E}_{\mu\nu}$ is the  effective non-local anisotropic stress-tensor; and the effective non-local energy flux on the brane, $\mathit{Q}_\mu = -\frac16\sigma h^{\;\uprho}_{\mu}\mathcal{E}_{\uprho\nu}u^\nu$, is originated from the bulk free gravitational field. Local corrections are encoded in the tensor \cite{GCGR}:	
\begin{eqnarray}\label{smunu}
S_{\mu\nu} = \frac{T}{3}T_{\mu\nu}-T_{\mu\kappa}T^\kappa_{\ \nu} + \frac{g_{\mu\nu}}{6} \Big[3T_{\kappa\tau}T^{\kappa\tau} - T^2\Big] \ ,
\end{eqnarray}
\noindent where $T_{\mu\nu}$ is the brane matter stress-tensor. Higher-order terms in Eq. (\ref{smunu}) are neglected, as the brane matter density
is negligible.  Denoting by $G_{\mu\nu}$ the Einstein tensor, the 4D effective Einstein's effective field equations read
\begin{equation}
G_{\mu\nu}
-T_{\mu\nu}-\mathcal{E}_{\mu\nu}(\sigma^{-1})-\frac{\sigma^{-1}}{4}S_{\mu\nu}=0 . \label{projeinstein}
\end{equation} 
Since $\mathcal{E}_{\mu\nu} \sim \sigma^{-1}$, it is straightforward to notice that in the infinitely rigid brane limit,  $\sigma \rightarrow \infty$, GR is recovered and the Einstein's field equations have the standard form  $G_{\mu\nu} = T_{\mu\nu}$. 

On the other hand, the AdS/CFT setup yields the effective
equations on the brane \cite{Shiromizu:2001jm,Randall:1999vf,Gubser:1999vj,deHaro:2000wj,Henningson:1998gx}:
\begin{eqnarray}
G_{\mu\nu}=8\pi G_4 T_{\mu\nu}+\frac{4}{{l\sqrt {|g|}}}
\left( \frac{\delta S{}
_{\rm ct}}{\delta g_{\mu\nu}} +
\frac{\delta \Gamma_{\rm CFT}}{\delta g_{\mu\nu}}  \right),
\label{19}
\end{eqnarray} where $l=4/K$ (here $K$ is the trace of the extrinsic curvature tensor) and $\Gamma_{\rm CFT}$ corresponds to the effective action of CFT in the
boundary, whose trace anomaly reads \cite{deHaro:2000wj,Henningson:1998gx}:
\begin{eqnarray}
g^{\mu\nu}\frac{\delta \Gamma_{\rm CFT}}{\delta g_{\mu\nu}}=
\frac{l^3}{16}
{\sqrt {|g|}}\Bigl(
R_{\mu\nu}
R^{\mu\nu}-{\textstyle \frac{1}{3}}
R^2  \Bigr), \label{eq:anomaly}
\end{eqnarray} where $R_{\mu\nu}$ and $R$ are the Ricci tensor and scalar of the four-dimensional metric. 
The quantity
$S_{\rm ct}{}$
 encodes $R^2$ terms of the counter-term, making the action finite, and $\delta S_{\rm ct}/\delta g_{\mu\nu}$ is traceless, 
\begin{eqnarray}
\frac{\delta S{}_{\rm ct}}{\delta g_{\mu\nu}}
&\! \approxeq &\! -\frac{l^3}{32}
\left[ \frac{1}{6}D_\mu D_\nu {}{}R\!-\!\frac{1}{2}\Box R_{\mu\nu}\!+\!\frac{1}{4}g_{\mu\nu}\left(\frac13\Box R\!+\!\frac13g_{\mu\nu}R^2\!-\!\frac{1}{4}R_{\alpha\beta}{}{}R^{\alpha\beta}
\right)\right.\nonumber\\
&&\left. +R^{\alpha\beta}R_{\mu\alpha\nu\beta}
\!-\!\frac{1}{3}RR_{\mu\nu}
\right].
\end{eqnarray}
Then, the trace part of Eq. (\ref{19}) reads $
R=-8\pi G_4 T-\frac{l^2}{4}
\Bigl(
R_{\mu\nu}
R^{\mu\nu}-{\textstyle \frac{1} {3}}
R^2  \Bigr)$.  Hence, in the linear order  the energy-momentum tensor of CFT is governed by the electric part of the Weyl tensor \cite{Randall:1999vf,Gubser:1999vj,deHaro:2000wj}:
\begin{eqnarray}
E^{\mu\nu} \approxeq - \frac{K}{{\sqrt {|g|}}}
\frac{\delta \Gamma_{\rm CFT}}{\delta g_{\mu\nu}}.
\end{eqnarray}
The effective Einstein's equations read 
\be
\label{5d4d}
R_{\mu\nu}-\frac12\, R\,g_{\mu\nu}
=
8\,\pi\,\gn\,T^{\rm eff}_{\mu\nu}
-\Lambda\,g_{\mu\nu}
\ ,
\ee
where $\gn= \lp/\mpl$, with $\mpl$ and $\lp$ the four-dimensional Planck mass and scale, respectively and 
$\Lambda$ is the cosmological constant, which will be neglected  hereafter. 
The effective stress tensor in Eq.~(\ref{5d4d}) contains the matter energy-momentum tensor on the brane,
the electric component of the Weyl tensor and the projection of the bulk energy-momentum tensor onto the
brane~\cite{maartens}. 
For static and spherically symmetric metrics,  
compact stellar distributions in 4D, which must be solutions of Eq. \eqref{projeinstein}, can be described  in Schwarzschild-like coordinates as
\begin{equation}
 ds^2 = -e^{\upnu(r)}\mathrm{d}t^2 + e^{\lambda(r)}\mathrm{d}r^2 + r^2\mathrm{d}\Omega^2 \ ,
 \label{eq:metric_general_spher_symm}
\end{equation}
The MGD provides a solution to Eqs.~\eqref{5d4d} by deforming the radial metric component of the corresponding
GR solution~\cite{covalle2,Ovalle:2014uwa}.
For the GR Schwarzschild metric, and dismissing terms of order $\sigma^{-2}$ or higher, one obtains~\cite{covalle2}
\begin{subequations}
\ba
\label{nu}
\textrm{e}^{\nu(r)}
&=&
1-\frac{2\,M}{r}
\,
\\
\textrm{e}^{-\lambda(r)}
&=&
\textrm{e}^{\nu(r)}\left[1+\frac{2\,{\ell}}{{2\,r-{3\,M}}}\right]
\ ,
\label{minus-lambda}
\ea
\end{subequations}
where $\ell\approxeq -\frac{1.352\left(1-\frac{3M}{2R}\right)}{\sigma R\left(1-\frac{2M}{R}\right)}$ is the length scale previously discussed in the Sect. \ref{intro}, being $M$ the ADM mass. In Eqs. \eqref{nu} and \eqref{minus-lambda} geometrized units, $\gn=c=1$, are adopted. There are two solutions of the equation $\textrm{e}^{-\lambda(r)}=0$, namely 
\begin{subequations}
\ba
\mathring{r}
&=&
2\,M\ ,
\\
r_-
&=&
\frac{3}{4}\,\mathring{r}-\ell
\ ,
\ea
\end{subequations}
so that $\mathring{r}>r_-$ for any $\ell>0$.
For studying the Hawking radiation, one is interested in the region outside $\mathring{r}$, that effectively acts as the event horizon,
and just note that $r_-$ is not a (Cauchy) horizon~\cite{covalle2}. 
\par
We just mention in passing that an explicit expression for $\ell$ in terms of $\sigma^{-1}$ can be obtained
by first considering a compact source of finite size $r_0$ and proper mass $M_0$~\cite{covalle2,Ovalle:2017fgl},
and then letting the radius $r_0$ decrease below $\mathring{r}$.
However, for practical purposes, it is more convenient and general to show the dependence on the length $\ell$.
For example, observational data impose bounds on the length $\ell$, from which bounds on $\sigma$ can be
straightforwardly inferred according to the underlying model~\cite{Casadio:2015jva,Casadio:2016aum}.  
The MGD and EMGD black holes  were respectively used in Refs. \cite{daRocha:2017cxu} and \cite{Fernandes-Silva:2018abr} to explore the observational signatures of SU($N$) dark glueball condensates and their gravitational waves.

A more general solution for the exterior radial metric component was derived in Ref. \cite{Casadio:2015gea}, under the extended minimal  geometric deformation, EMGD, with 
\begin{equation}
e^\upnu = \left ( 1 - \frac{2M}{r}\right )^{k+1} \ ,
  \label{eq:temp_emgd}
\end{equation}
\noindent where $k$ is a constant known as the exponential deformation parameter. Naturally, $k=0$ results no temporal geometric deformation, being directly associated with the Schwarzschild metric when $\sigma\to\infty$. For $k=1$, one has \cite{Casadio:2015gea}
\begin{eqnarray}
e^{\upnu(r)} &=& 1 - \frac{4M}{r} + \frac{4M^2}{r^2} \ ,\nonumber\\
 e^{-\lambda(r)} &=& 1 - \frac{2M - \kappa_1}{r} + \frac{2M^2 - \kappa_1 M}{r^2} \ ,
\end{eqnarray}
\noindent for $\kappa_1 = \dfrac{M\chi}{1 - M/R}$.
Now, in order to the radial metric component asymptotically approach the Schwarzschild behavior with ADM mass $\mathbb{M}_1 = 2M$, $
 e^{-\lambda(r)} \sim 1 - \frac{2\mathbb{M}_1}{r} + \mathcal{O}(r^{-2})$, one  must necessarily have $\kappa_1 = -2M$. In this case, the temporal and spatial components of the metric will be inversely equal to each other (as it is the case of the Schwarzschild solution), containing a tidal charge $\mathbb{Q}_1 = 4M^2$ reproducing a  solution that is  tidally charged by the Weyl fluid \cite{dadhich}:
\begin{equation}
 e^\upnu = e^{-\lambda} = 1 - \frac{2\mathbb{M}_1}{r} + \frac{\mathbb{Q}_1}{r^2}
 \label{eq:k=1}
\end{equation}
It is worth to emphasize that the metric of Eq. (\ref{eq:k=1}) has a degenerate event horizon at $r_h = 2M = \mathbb{M}_1$. Since the degenerate horizon lies behind the Schwarzschild event horizon, $r_h = \mathbb{M}_1 < r_s = 2\mathbb{M}_1$, bulk effects are then responsible for decreasing the gravitational field strength on the brane.

Now the exterior solution for $k=2$ can be constructed, making Eq. (\ref{eq:temp_emgd}) to yield 
\begin{equation}\label{eq:k=2nu}
 e^{\upnu(r)}  = 1 - \frac{2 \mathbb{M}_2}{r} +\frac{\mathbb{Q}_2}{r^2} - \frac{2\mathbb{Q}_2 \mathbb{M}_2 }{9r^3}\ ,
\end{equation}
where $\mathbb{Q}_2 = 12M^2$ and $\mathbb{M}_2 = 3M$. 
The radial component, on the other hand, reads
\begin{equation}\label{eq:k=2lambda-minus}
  e^{-\lambda(r)}  = \frac{1}{ 1 - \frac{2 \mathbb{M}_2}{3r}} \sum_{m=0}^8\frac{c_m}{r^m}~,
\end{equation}
where the coefficients $c_m\equiv c_m(\mathbb{M}_2,\mathbb{Q}_2,\textsc{s})$ are
\begin{subequations}\label{k=2Coeffs}
\begin{eqnarray}
\!\!\!\!\!\!\!\!\!\!\!\!\!\!\!\!\!\!\!\!\!\!\!\!\!\!\!\!\!\!\!\!c_0&=&1~,~\quad\qquad\quad\quad\quad\quad
c_1=\textsc{s}-\frac{4 \mathbb{M}_2}{3}~,\quad \quad\quad\quad
c_2=\frac{1}{6}\left(5\mathbb{Q}_2-7\textsc{s}\mathbb{M}_2\right)~,\\
\!\!\!\!\!\!\!\!c_3&\!=\!&\frac{\mathbb{M}_2}{12}  (7 \textsc{s} \mathbb{M}_2\!-\!5 \mathbb{Q}_2)~,~
c_4\!=\!\frac{25\mathbb{Q}_2^2}{288}\!-\!\frac{7}{216}\textsc{s} \mathbb{M}_2^3~,~
c_5\!=\!\frac{35}{1296} \textsc{s} \mathbb{M}_2^4\!-\!\frac{35}{1728} \mathbb{Q}_2^2 \mathbb{M}_2,\\
\!\!\!\!\!\!\!\!c_6&=&\frac{5 \mathbb{Q}_2^3}{20736}-\frac{7 \textsc{s} \mathbb{M}_2^5}{2592}~,~\quad
c_7=\frac{28 \textsc{s} \mathbb{M}_2^6-15 \mathbb{Q}_2^3 \mathbb{M}_2}{186624}~,~
c_8=\frac{5 \mathbb{Q}_2^4}{4644864}-\frac{\textsc{s} \mathbb{M}_2^7}{279936},
\end{eqnarray}
\end{subequations}
and $\textsc{s}=R\chi\left(1-2\mathbb{M}_2/3R\right)/\left(2-\mathbb{M}_2/3R\right)^7$~. The asymptotic Schwarzschild behavior is then assured when $ \textsc{s}= -\mathbb{M}_2/96$. In this case, the degenerate event horizon is at $r_e \approx 1.12 \mathbb{M}_2$ \cite{Casadio:2015gea}. 
Hence, the bulk Weyl fluid weakens gravitational field effects.
The classical tests of GR applied to the EMGD metric provide the following constraints on the value of the deformation parameter, $k \lesssim 4.2$ for the gravitational redshift of light. 
The standard MGD  corresponds to $k=0$, whereas the Reissner--Nordstr\"om 
solution represents the $k=1$ case with the ADM mass $\mathbb{M}_1$, instead.

\section{HEE in MGD spacetimes}\label{sect:MGD}

\begin{figure}[hbtp!]
    \centering
    \includegraphics[scale=0.5]{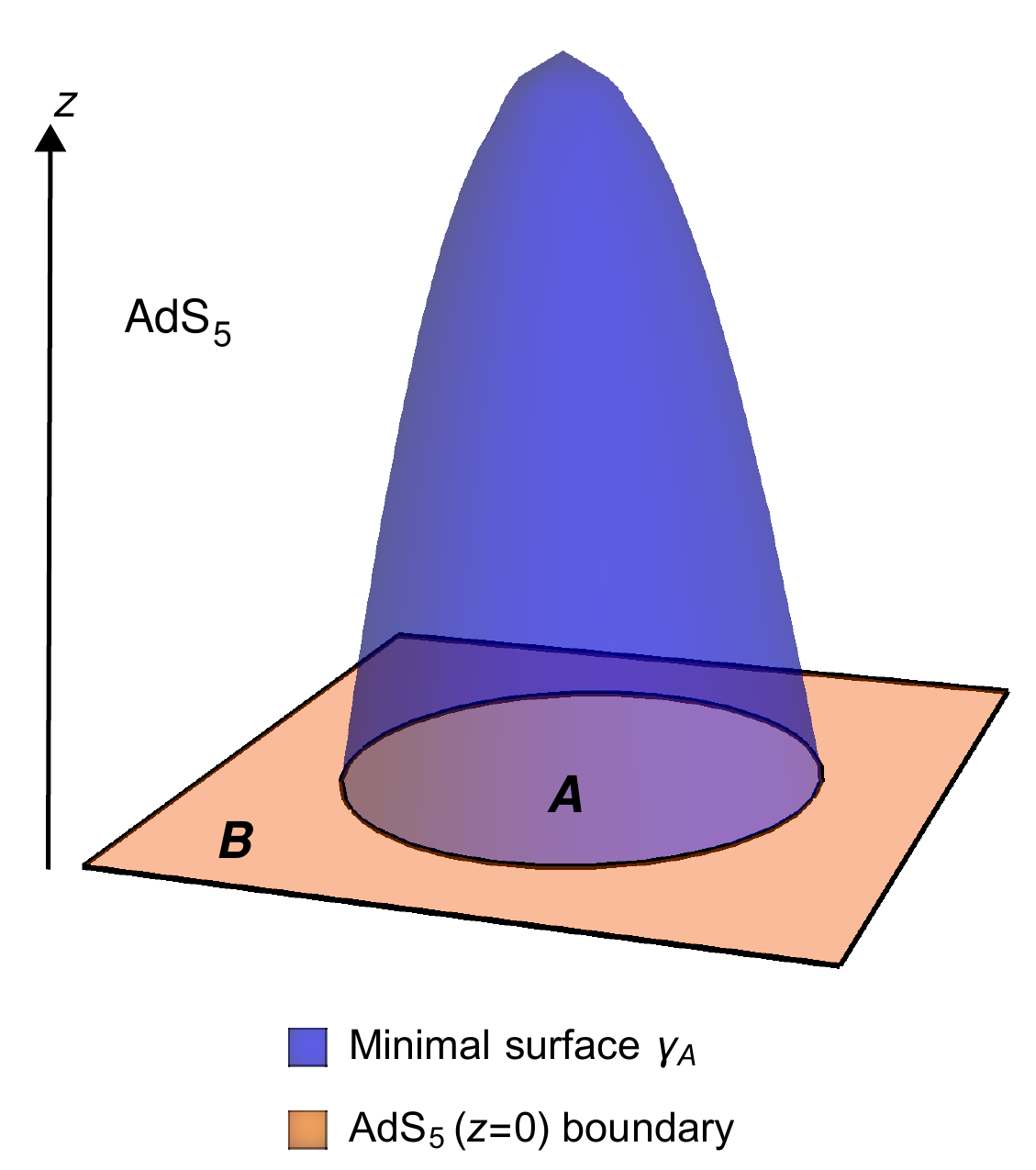}
    \caption{Ryu and Takayanagi prescription of the HEE. The light blue codimension 2 minimal surface $\gamma_A$, anchored on the boundary $\partial A$ of the entangling region $A$ in the AdS${}_5$ boundary,  has hypersurface area determining the EE related to the region $A$.
}
    \label{fig:fig}
\end{figure}

\quad\; The EE $S_A$ in QFTs represents the von Neumann entropy of the reduced density matrix, when one spreads degrees of freedom inside a 3D spacelike submanifold $B$ in a given 4D QFT, which is a complement of a manifold $A$.
$S_A$ ia responsible to quantify the correlation between $A$ and $B$, seen as two physical subsystems. In other words, $S_A$ corresponds to the entropy observed in $A$, by an observer that has no access to
$B$. The EE does not vanish at the zero temperature limit (Fig.~\ref{fig:fig}).
As the amount of information in the subsystem $B$ can be computed by the EE $S_A$, one may argue which component of the 
AdS$_5$ bulk is in charge for computing $S_A$ in the dual gravity.

The definition of EE can be implemented, once one considers QFTs \cite{Ryu:2006bv}. At zero temperature, the quantum system is described by the
pure ground state $|\Psi\lb$. Then, the density
matrix is that of the pure state $ \rho_{tot}=|\Psi\rangle \langle
\Psi|.$ The von Neumann entropy of the total system
is clearly zero $ S_{tot}= -\mathrm{tr}\, \rho_{tot} \log
\rho_{tot}=0$. Splitting the total system into two subsystems $A$ and $B$, the observer that has access only to the subsystem $A$ will feel as
if the total
system is described by the reduced density matrix $\rho_A= \mathrm{tr}_{B}~\rho_{tot}.$ 
Now one defines the EE of the
subsystem $A$ as
the von Neumann entropy of the reduced
density matrix $\rho_A$, namely, 
$S_A =
- \mathrm{tr}_{A}\,
\rho_{A} \log \rho_{A}$. 
If the density matrix $\rho_{tot}$ is pure, then
as $B$ is the complement of $A$, it follows that 
$ S_A=S_B$.   This equality is violated
              at finite temperature. 
One can find the subadditivity relation, $
 S_{A+B}\le S_{A}+S_{B}.$

 More precisely, considering a QFT on a 4D spacetime splitting, 
$\mathbb{R}\times \Sigma_3$, into timelike vector field and a 3D
 spacelike manifold, $\Sigma_3$.
Define a 3D submanifold $B\subset
\Sigma_3$ at fixed time $t=t_0$ as the complement of 
$A$ with respect to $\Sigma_3$. The boundary $\de A$ of $A$, divides the
manifold $\Sigma_3$ into two complementary submanifolds $A$ and $B$.  As the EE diverges in the 
continuum limit, an UV cutoff $a$ is needed.
Then the coefficient of the divergence is proportional to the area of
the boundary $\de A$, \be S_A\approx\alpha\cdot \frac{\mbox{Area}(\de
A)}{a^{2}},
 \label{divarea}\ee
where $\alpha$ is a constant. 
Employing the
Poincar\'e metric of AdS$_5$ with radius $R$,
\be ds^2=\frac{R^2}{z^2}\left(dz^2-dx_0^2+
dx_idx^i\right), 
\label{Poincare} \ee
the dual CFT$_{4}$ is supposed to live on the boundary of AdS$_{5}$
which is $R^{1,3}$
at $z\to 0$ spanned by the coordinates $(x^0,x^i)$. The bulk conformal  coordinate
$z$ in AdS$_{5}$ is interpreted as the length scale of the dual CFT$_{4}$.
Since the metric diverges
in the limit $z\to 0$, we put a cutoff by imposing $z \geq a$. Then the boundary is situated
at $z=a$.

Although AdS/CFT is based on an AdS spacetime (\ref{Poincare}), it can be also used to any asymptotically AdS$_5$ spacetime, encompassing AdS black branes. 
Now we are in a position to present how to calculate the entanglement
entropy in CFT$_{4}$ from the gravity on AdS$_{5}$. 
 In the setup (\ref{Poincare}), one extends $\partial A$ to a surface $\gamma_A$, such
that $\partial \gamma_A=\partial A$. One has to choose the minimal area surface among them. 
In this setup the EE $S_A$ in CFT$_{4}$ can be computed \cite{Ryu:2006bv,Hubeny:2007xt,Emparan:2006ni}. 
\begin{equation} S_{A}=\frac{{\rm Area}(\gamma_{A})}{4G_{5}} \ .
\label{eq:HEE}
\end{equation}
 To choose the minimal
surface as in (\ref{eq:HEE}) means that one defines the severest
entropy bound \cite{Nishioka:2009un} so that it has a chance to
saturate the bound.

There is an identification of the 4D entanglement entropy QFT with a certain geometrical quantity in 5D gravity, then generalizing the black hole entropy.  In the particular case of the membrane paradigm, this identification implements the relationship between black hole entropy and entanglement entropy in the induced gravity setup \cite{Nishioka:2009un}.

We will study the HEE from two perspectives: the MGD, in this section, and the EMGD solutions, in the next one. For both of them, one needs to understand how the first law of HEE holds in the context of the membrane paradigm. The dual theory can be defined on a boundary located at two kind of distance ranges: \emph{$($i$)$: far from the horizon} -- a finite large radial coordinate denoted by  $r_\infty$, and \emph{$($ii$)$: almost on the horizon} -- a small displacement from the horizon, named $\delta r\equiv r-\mathring{r}$, where $\mathring{r}$ is the horizon \textit{situs} on spacetime. The  MGD  HEE will be implemented under these perspectives and scrutinized in what follows. The metric in Eq.~\eqref{eq:metric_general_spher_symm} is employed, where the temporal and radial components are respectively set by Eqs. (\ref{nu}) and (\ref{minus-lambda}).

\subsection{Far from the horizon}\label{subsect:MGDbeyond}

In the region far from the horizon, the boundary manifold is placed at $r=r_\infty$ that is far away from the event horizon. Let one considers a circle, in spherical coordinates defined by the azimuthal angle $\theta=\theta_0$, responsible to enclose the entangling surface. The radial coordinate function, $r=r(\theta)$, describes the minimal surface whose boundary is the entanglement surface. In addition, the minimization of the area function, 
\begin{equation}\label{eq:area}
\mbox{Area}(\gamma_A)={2\pi}\int_0^{\theta_0}\mathrm{d}\theta\left\{r\sin\theta
\left[\textrm{e}^{\lambda(r)}\left(\frac{\mathrm{d}r}{\mathrm{d}\theta}\right)^{2}+r^2\right]^{1/2}\right\},
\end{equation}
with boundary condition $r(\theta_0)=r_\infty$, plays a prominent role in  computing the minimal surface. Obtaining the global minimum of the area yields the HEE, by employing Eq. \eqref{eq:HEE}. Eq. \eqref{eq:area} reads 
\begin{equation}\label{eq:Lag1}
\mbox{Area}(\gamma_A)=\int_{y_0}^1 \mathrm{d}y \, \mathcal{L}^{\scalebox{.5}{MGD}}~, \end{equation}
where $\mathcal{L}^{\scalebox{.5}{MGD}} = 2\pi r\left[(1-y^2)\mathcal{F}\dot{r}^2+r^2\right]^{1/2}$, $y=\cos\theta$ and $y_0=\cos\theta_0$. The dot designates the derivative with respect to $y$ and $\mathcal{F}=\mathcal{F}\left(r(y)\right)\equiv\textrm{e}^{\lambda\left(r(y)\right)}$. 
Applying the variational method, one varies Eq.~\eqref{eq:Lag1} with respect to $r(y)$, yielding the following ODE: 
\begin{equation}\label{eq:EOM}
(y^2-1)\left[2\mathcal{F}r^2\ddot{r}-2y\mathcal{F}^2\dot{r}^3+\left(r\frac{\mathrm{d}\mathcal{F}}{\mathrm{d}r}-6\mathcal{F}\right)r\dot{r}^2\right]
+4y\mathcal{F}r^2\dot{r}+4r^3=0~.
\end{equation}
 Eq.~\eqref{eq:EOM} is strongly nonlinear. Therefore, a way to attenuate it is to  attribute $\mathcal{F}\equiv\mathcal{F}\left(r(y)\right)=1$, to yield  $r= w_0/y$ as the simplest solution to be achieved.
In addition, according to Ref. \cite{Sun:2016dch}, one can derive  nontrivial solutions of Eq.~\eqref{eq:EOM}, working with series expansions, respectively for $\mathcal{F}\left(r(y)\right)$ and $r(y)$: 
\begin{subequations}
\begin{eqnarray}
\mathcal{F}\left(r(y)\right)&=&1-\sum_{j=1}^{\infty} g_j(y)\varepsilon^j~,\label{eq:Fexp}\\
r(y)&=&\frac{w_0}{y}+\sum_{j=1}^{\infty}r_j(y)\varepsilon^j~.\label{eq:rexp}
\end{eqnarray}
\end{subequations}
Here $\varepsilon$ denotes a small dimensionless parameter,  relating the black hole mass $M$ to $r_\infty$ by $\varepsilon = \frac{M}{r_\infty}.$ The $\mathcal{O}(\varepsilon)$ terms in the expansions \eqref{eq:Fexp} and \eqref{eq:rexp} may indicate corrections 
regarding the black hole collapse itself. It is worth to emphasize that the $0^{\rm th}$-order term, $r(y)=w_0/y$, in \eqref{eq:rexp} is the solution corresponding to $\mathcal{F}=1$.  

Now, considering the $\mathcal{F}$ function for the MGD spacetime, encoded in Eq. \eqref{minus-lambda}, one finds, up to the $2^{\rm nd}$-order in the $g_j(y)$ functions in the series \eqref{eq:Fexp},  
\begin{subequations}\label{g-functions}
\begin{eqnarray}
g_1(y)&=&\frac{(\xi-2)y r_\infty}{w_0}~,\\
g_2(y)&=&\frac{y^2 r_\infty}{2w_0^2}\left[r_\infty(-8+7\xi-2\xi^2)+2(\xi-2)r_1(y)\right],
\end{eqnarray}
\end{subequations}
where, due to dimensional analysis, the MGD parameter related to the expansion parameter can be written as $\ell=\xi M.$  Higher order terms in Eq.~\eqref{eq:Fexp} can be forthwith derived. 
The set of auxiliary functions $\{g_1(y), g_2(y),\ldots\}$ in Eq. \eqref{g-functions} is important to solve Eq. \eqref{eq:EOM} order  by order \cite{Sun:2016dch,Kim:2015rvu}. We intend here to pursuit the possible modifications to the HEE up to the $2^{\rm nd}$-order. Hence, the calculation of the $r$-functions immediately follows, which are necessary to provide the HEE corrections up to $2^{\rm nd}$-order.

The $1^{\rm st}$-order ODE, taking $1^{\rm st}$-order terms in $\varepsilon$, reads
\begin{equation}\label{r1EDOMGD}
\ddot{r}_1(y) + \frac{\left(5 y^2-3\right)}{y \left(y^2-1\right)}\dot{r}_1(y)+\frac{\left(3 y^2-1\right)}{y^2\left(y^2-1\right)}r_1(y)=\frac{\left(3 y^2+1\right) (2-\xi)r_{\infty }}{y^2 \left(y^2-1\right)}~.
\end{equation}
Eq.~\eqref{r1EDOMGD} carries $\textsc{D}_1$ and $\textsc{D}_2$ as constants of integration, whose values are determined by the finiteness condition. Hence, to avoid divergences at $y=1$, as $y=\cos\theta\in\left[\cos\theta_0,1\right]$, one needs to set $\textsc{D}_2=(2-\xi)r_\infty$. Besides, using the boundary condition $r_1(y)=0$ yields $\textsc{D}_1=(\xi-2)r_\infty\{y_0+2\log[y_0/(1+y_0)]\}$. Therefore, the first $r$-function reads
\begin{equation}\label{r1MGD}
    r_1(y)=\frac{(2-\xi)r_\infty}{2y}\left[y-y_0-2\log\left(\frac{1+y}{1+y_0}\right) +2\log\left(\frac{y}{y_0}\right)\right]~.
\end{equation}
Importantly, there is a subtle restriction due to limitations in the perturbative expansion, as aforementioned in Ref. \cite{Sun:2016dch}. In fact,  the $y=0$ point is never reached. Hence, the validity of the solution $r_1(y)$ is contained in the interval $\theta_0<\pi/2$ or, equivalently, $y\in (0,1)$.

Going to the $2^{\rm nd}$-order in $\varepsilon$, and employing the $r_1(y)$ solution in Eq. \eqref{r1MGD}, yields
\begin{equation}\label{r2EDOMGD}
    \ddot{r}_2(y)+\frac{\left(5 y^2-3\right)}{y\left(y^2-1\right)} \dot{r}_2(y)+\frac{\left(3 y^2-1\right)}{y^2 \left(y^2-1\right)}r_2(y) = \mathcal{P}(y)~.
\end{equation}
with
\begin{equation}
 \mathcal{P}(y) = \frac{r_\infty^2}{2w_0}\left[\frac{(\xi -2)^2(y^3+3y-4)+2\xi y^3}{y^2 \left(y^2-1\right)}\right].
\end{equation}
Proceeding analogously as in the solution of Eq. \eqref{r1EDOMGD} implies that 
\begin{eqnarray}\label{r2pre}
 \!\!\!\! \!\!\!\! \!\!\!\!r_2(y)=\frac{\textsc{D}_3}{y}+\frac{r_\infty^2}{16 w_0 y}\left[\textsc{H}_1(\xi)y^2\!-\!\textsc{H}_2(\xi)\log(1\!-\!y)\!+\!\textsc{H}_3(\xi)\log(1\!+\!y)\right]+\frac{\textsc{D}_4\left[2\log y\!-\!\log(1\!-\!y^2)\right]}{2y},
\end{eqnarray}
where $\textsc{H}_1(\xi)=(\xi-2)^2+2\xi$,~$\textsc{H}_2(\xi)=36-38\xi+9\xi^2$, and $\textsc{H}_3(\xi)=92 - 90\xi + 23\xi^2$. Once more,  computing of $\textsc{D}_4$ and $\textsc{D}_3$ requires the preclusion of divergences at $y=1$ and the boundary condition $r_2(y_0)=0$, respectively. With this setup, they read 
\begin{subequations}
\begin{eqnarray}
\textsc{D}_4&=&-\frac{r_\infty^2}{8w_0}\textsc{H}_2(\xi)~,\\
\textsc{D}_3&=&-\frac{r_\infty^2}{16 w_0}\left[\textsc{H}_1(\xi) y_0^2 - 2\textsc{H}_2(\xi)\log(y_0) + \textsc{J}(\xi)\log\left(1+y_0\right)\right]~,
\end{eqnarray}
\end{subequations}
with $\textsc{J}(\xi)=32(\xi-2)^2$. Hence, the complete form of the second $r$-function is given by 
\begin{equation}
    r_2(y)=\frac{r_\infty^2}{16 w_0 y}\left[\textsc{H}_1(\xi)(y^2-y_0^2)-2\textsc{H}_2(\xi)\log\left(\frac{y}{y_0}\right)+\textsc{J}(\xi)\log\left(\frac{1+y}{1+y_0}\right)\right]~.
\end{equation}
As the last step, we proceed to the expansion $\mathcal{L}^{\scalebox{.5}{MGD}}=\mathcal{L}_0^{\scalebox{.5}{MGD}}+\varepsilon\mathcal{L}_1^{\scalebox{.5}{MGD}}+\varepsilon^2\mathcal{L}_2^{\scalebox{.5}{MGD}}+\cdots$, within the formula for the area shown in Eq. \eqref{eq:Lag1}. From now, as formely mentioned, the $r$-functions are employed to compute each order of the contribution for the HEE, $\mathcal{S}^{\scalebox{.5}{MGD}}=\mathcal{S}_0+\mathcal{S}_1^{\scalebox{.5}{MGD}}+\mathcal{S}_2^{\scalebox{.5}{MGD}}+\cdots$. Besides this expansion will be considered, including terms of $2^{\rm nd}$-order. Next, the detailed computation of each order is provided.

For the $0^{\rm th}$-order, one has the following expression:
\begin{equation}\label{S0MGDbeyond}
    \mathcal{S}_0^{\scalebox{.5}{MGD}}=\frac{A_0}{4}=\frac{1}{4}\int_{y_0}^0dy\mathcal{L}_0=\int_{y_0}^0dy \frac{2 \pi  w_0^2}{y^3}=\frac{1}{4} \pi  w_0^2 \left(\frac{1}{y_0^2}-1\right),
    \end{equation}
 whereas  the $1^{\rm st}$-order reads 
\begin{eqnarray}\label{S1MGDbeyond}
    \mathcal{S}_1^{\scalebox{.5}{MGD}}&=&\frac{A_1}{4}=\frac{\varepsilon}{4}\int_{y_0}^0dy\mathcal{L}_1=\frac{(2-\xi)}{4}\pi r_\infty M(1-y_0)^2.
\end{eqnarray}
 Compared with the results obtained in Ref. \cite{Sun:2016dch}, our results show an interesting novelty. Although the $0^{\rm th}$-order term of the entanglement entropy remains the same, the $1^{\rm st}$-order corrections for the HEE display the MGD parameter, $\xi$,  which carries the signature of the finite brane tension, within this order of correction, into the HEE. The general relativistic limit, $\sigma\to\infty$, yields $\xi\to 0$, recovering the $1^{\rm st}$-order correction to the HEE in Schwarzschild spacetime. Besides, the $0^{\rm th}$-order of the entropy is proportional to $r_\infty^2$, since $w_0=r_\infty\cos\theta_0$, whereas the $1^{\rm st}$-order one is proportional to $r_\infty$, with the MGD parameter increasing the numerical factor. This indicates a small contribution of the $1^{\rm st}$-order, compared to the $0^{\rm th}$-order --  as pointed out in \cite{Sun:2016dch} -- even in the presence of the MGD parameter $\xi$.

To analyze the signature of the MGD parameter on the correction, at a given order, in the HEE, a new  quantifier can be introduced. We define the $n^{\rm th}$-order corrections ratio as
\begin{equation}\label{Phi_n}
    \Phi^{\scalebox{.5}{MGD}}_n=\frac{\mathcal{S}_ n^{\scalebox{.5}{MGD}}}{S_n^{\scalebox{.6}{Schw}}}~,
\end{equation}
where $\mathcal{S}_ n^{\scalebox{.5}{MGD}}$ and $S_n^{\scalebox{.5}{Schw}}$ are the $n^{\rm th}$-order corrections to the HEE in MGD and Schwarzschild spacetimes, respectively. 
Hence, one has $\Phi^{\scalebox{.5}{MGD}}_0=\mathcal{S}_ 0^{\scalebox{.5}{MGD}}/\mathcal{S}_ 0^{\scalebox{.5}{Schw}}=1$, as  the $0^{\rm th}$-order corrections are equal. Meanwhile,  the $1^{\rm th}$-order corrections yield 
\begin{equation}\label{phi1MGD}
    \Phi^{\scalebox{.5}{MGD}}_1=\frac{\mathcal{S}_ 1^{\scalebox{.5}{MGD}}}{S_1^{\scalebox{.5}{Schw}}}=1-\frac{\xi}{2}~.
\end{equation}
As $\xi=\ell/M$ and $\ell<0$, then both $\mathcal{S}_ 1^{\scalebox{.5}{MGD}}$ and $\mathcal{S}_ 1^{\scalebox{.5}{Schw}}$ are positive, representing, at this order of correction, a linear increment of the EE depending on the MGD parameter.

Now, the next order reads 
\begin{eqnarray}\label{S2MGDbeyond}
    \mathcal{S}_2^{\scalebox{.5}{MGD}}&=&\frac{A_2}{4}\nonumber\\&=&\frac{\varepsilon^2}{4}\int_{y_0}^0dy\mathcal{L}_2=\frac{\pi M^2}{32} \left[\textsc{U}_1(\xi,y_0)+\textsc{U}_2(\xi)\log \left(\frac{2}{1+y_0}\right)+\textsc{U}_3(\xi)\log(y_0)\right]~,\nonumber
\end{eqnarray}
with ancillary functions $\textsc{U}_1(\xi,y_0)=\left[2\xi(13\!-\!3 y_0)\!-\!(\xi^2\!+\!4)(7\!-\!y_0)\right](1-y_0)$, $\textsc{U}_2(\xi)=16(\xi-2)^2$ and $\textsc{U}_3(\xi)=2\left[(\xi-2)^2-2\xi\right]$.
One can notice the contribution of the MGD parameter,  encoding  the finite brane tension, as one compares with the HEE for the Schwarzschild spacetime, corresponding to $\ell\to 0$ and, hence,  $\xi\to 0$. Henceforth, in the general relativistic case of a rigid brane, $\sigma\to\infty$, one recovers the $2^{\rm nd}$-order correction for Schwarzschild spacetimes. On the other hand, the $2^{\rm nd}$-order corrections ratio are given by 
\begin{equation}\label{phi2MGD}
    \Phi_2^{\scalebox{.5}{MGD}}=\frac{\mathcal{S}_ 2^{\scalebox{.5}{MGD}}}{S_2^{\scalebox{.5}{Schw}}}=1 + \frac{\xi}{4}\left(\xi-6\right)+4\xi\left[\frac{ 1-y_0-2 \log\left(\frac{2}{1+y_0}\right)}{7-8 y_0+y_0^2-2\log y_0-16 \log\left(\frac{2}{1+y_0}\right)}\right]~.
\end{equation}
Both corrections, the $1^{\rm st}$- and the $2^{\rm nd}$-order ones, have the MGD parameter as a dominant variable, when considering  the minimal surface in large range, correspondly, the lower limit very close to zero. The $1^{\rm st}$-order ratio  does not depend on such range. However, the $2^{\rm nd}$-order ratio has the limit
\begin{equation}\label{S2limitMGD}
    \Phi_2^{\scalebox{.5}{MGD}}|_{y_0\to 0}=1+\frac{\xi}{4}(\xi-6)~.
\end{equation}
As $\xi<0$, it is observed an increment of the value of this order of correction to the HEE. Irrespectively of the limit taken, the limit $\xi\to0$ recovers the $2^{\rm nd}$-order correction for the HEE in a Schwarzschild spacetime.

\begin{figure}[hbtp!]
    \centering
    \includegraphics[scale=0.5]{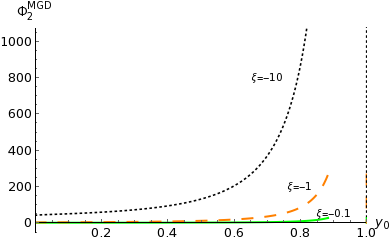}
    \caption{Ratio according with the brane tension and the range of the boundary.}
    \label{fig:phi2MGD}
\end{figure}

In general, the ratio depends on the finite brane tension and the lower limit of the minimal area. Fig.~\ref{fig:phi2MGD} displays such behavior. It is particularly important to notice that, since $\xi < 0$, a decrement of such contribution is observed, providing another relevant signature of the MGD parameter. Here, lower values of the brane tension contribute to diminish the HEE in MGD black holes.

By completeness, let us examine a restriction on $\xi$ to obtain the $2^{\rm nd}$-order contribution to the HEE in both MGD and Schwarzschild spacetimes. In such situation, the equality  $\Phi_2^{\scalebox{.5}{MGD}}=1$ holds, whenever the terms on the rhs of Eq. \eqref{phi2MGD} equal to 1. Let us denote the values of $\xi$ (eventually dependent on $y_0$) that satisfy this condition by $\xi_0$. Looking at Eq. \eqref{phi2MGD}, there are two solutions: the trivial one, $\xi_0=0$, and 
\begin{equation}
    \xi_0(y_0) = 6 + 16\left[\frac{y_0-1 + 2\log \left(\frac{2}{1+y_0}\right)}{y_0^2-8 y_0+7+2 \log y_0-16 \log \left(\frac{2}{1+y_0}\right)}\right]~.\label{ffff}
\end{equation} This result is quite relevant. In fact, the MGD parameter could produce an equal  correction ratio, depending on the lower limit of integration to compute the minimal area. However, as  $\xi<0$, such an exclusive value is not allowed, due to the fact that $\xi_0(y_0)>0$, for any value of $y_0$ in $(0,1)$.

\begin{figure}[htb]
    \centering
    \includegraphics[scale=0.5]{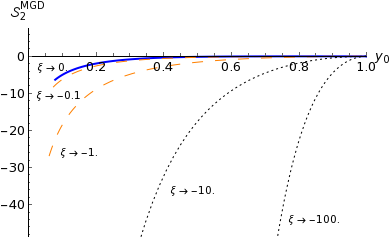}
    \caption{$S_2$ for MGD spacetimes for specific values of the brane tension and related to the lower limit of integration $y_0$.}
    \label{fig:S2MGD}
\end{figure}

\begin{figure}[htb]
    \centering
    \includegraphics[scale=0.5]{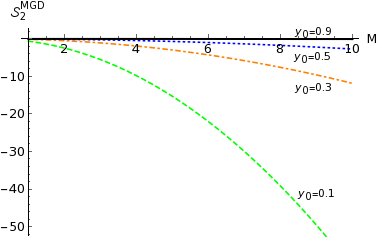}
    \includegraphics[scale=0.5]{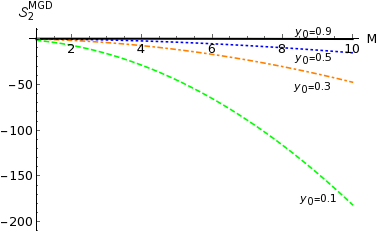}
    \includegraphics[scale=0.5]{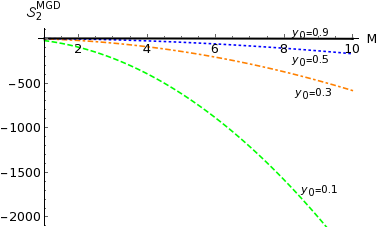}
    \includegraphics[scale=0.5]{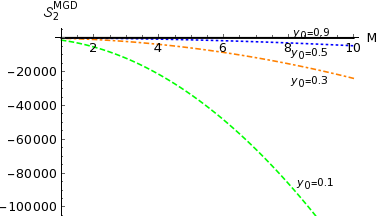}
    \caption{The $2^{\rm nd}$-order corrections for MGD spacetimes for $\xi=-0.1$, $\xi=-1$, $\xi=-10$ and $\xi=-100$ -- from the top to the bottom, from the left to the right, respectively -- varying the  mass parameter.}
    \label{fig:S2MGD_M_and_xi}
\end{figure}

\begin{figure}[htbp!]
\begin{center}
\includegraphics[scale=0.5]{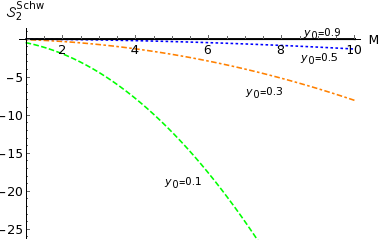}
\end{center}
\caption{Profile of the HEE $2^{\rm nd}$-order corrections in Schwarzschild black hole related to the mass parameter $M$, for  distinct values of $y_0$.} \label{fig:S2Schwarz}
\end{figure}

Besides, Fig. \ref{fig:S2MGD} displays the behavior of the $2^{\rm nd}$-order correction to the HEE in MGD spacetimes. It shows that the order of the correction in MGD spacetime is always negative and more intense than the same order of correction in Schwarzschild spacetime. This fact could be noticed by realizing the positivity of the ratio between both of them.
Fig.~\ref{fig:S2MGD_M_and_xi} considers the $2^{\rm nd}$-order correction for the MGD spacetime, by fixing $\xi$ and $y_0$ to different values of the black hole mass.
 For comparison, Fig.~\ref{fig:S2Schwarz} displays the increment of the $2^{\rm nd}$-order correction in a Schwarzschild black hole, as a function of the mass.

\begin{figure}[htb]
    \centering
    \includegraphics[scale=0.5]{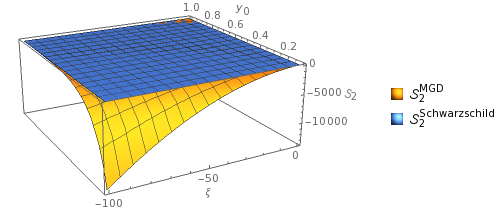}
    \caption{The behavior of the HEE $2^{\rm nd}$-order corrections in both MGD and Schwarzschild spacetimes.}
    \label{fig:S2MGD_M_y0}
\end{figure}

 One can notice the increment of this order of correction as  the black hole mass increases and, simultaneously, the decrement of $y_0$, which contributes with the extension of the minimal area. The smaller the brane tension, the greater the magnitude of correction in this order is, even with a minimal surface of small size. Besides,  Fig.~\ref{fig:S2MGD_M_y0} illustrates the behavior of the HEE $2^{\rm nd}$-order corrections in both MGD and Schwarzschild spacetimes, 
whereas the minimal surface size is a function of the black hole mass, $M$. 
\newpage

A small value of the brane  tension contributes to the increment of the HEE $2^{\rm nd}$-order correction in a MGD spacetime more intensely than the same correction in Schwarzschild spacetimes. The surface representing the HEE $2^{\rm nd}$-order correction in Schwarzschild spacetimes has an almost steady declination, when compared to the declination to the HEE $2^{\rm nd}$-order correction in a MGD spacetime.

Finally, one can notice the first law of HEE, as $\delta\mathcal{S}=\mathcal{S}-\mathcal{S}_0\propto M$, regarding a vast range of the brane tension, within precise phenomenological bounds \cite{Casadio:2016aum,Fernandes-Silva:2019fez}.

\subsection{Almost on the horizon}\label{subsect:MGDalmost}
\par 
\quad\; Inspired and motivated by Refs. \cite{Sun:2016dch,Carlip:1998wz}, the MGD black hole entropy, underlying the almost on the horizon boundary will be analyzed, using Eqs. (\ref{nu}, \ref{minus-lambda}). To simplify, the notations $\mathring{r}=2M$ and $r=\uprho^2+\mathring{r}$ makes implicit that $\uprho>0$ and $r>0$. Clearly, the event horizon is located at $\uprho=0$. Hence,
\begin{equation}
\mathrm{d}s^2=\left(\frac{r-\mathring{r}}{r}\right)\mathrm{d}t^2 + \left(\frac{r}{r-\mathring{r}}\right)\left[1+\frac{\ell}{\left(r-\frac{3}{4}\mathring{r}\right)}\right]^{-1}\mathrm{d}r^2
+r^2(\mathrm{d}\theta^2+\sin^2\theta \mathrm{d}\varphi^2).    
\end{equation}
 One sets a boundary almost on the horizon considering $\uprho_0=\varepsilon\sqrt{\mathring{r}}$, where $\varepsilon\ll 1$. The entangling surface is shaped as the $\theta=\theta_0$ circumference. Such a configuration yields an induced metric on the $t$-constant manifold, described by
\begin{equation}
\mathrm{d}\hat{s}^2=\left[4h(\uprho)\left(
\frac{\mathrm{d}\uprho}{\mathrm{d}\theta}\right)^2+g(\uprho)^2\right]\mathrm{d}\theta^2
+[g(\uprho)\sin\theta]^2\mathrm{d}\varphi^2,
\end{equation}
where $h(\uprho)=g(\uprho)f(\uprho)$, $g(\uprho)=\uprho^2+\mathring{r}$ and
$
    f(\uprho)=\left[1+\frac{\ell}{\left(\uprho^2+\frac{1}{4}\mathring{r}\right)}\right]^{-1}\!\!\!,$ 
with $\uprho\equiv\uprho(\theta)$.

Finding $\uprho$ means to minimize the surface area
\begin{equation}\label{area-min}
A=\int_{y_0}^1 \mathrm{d}y ~\tilde{\mathcal{L}}^{\scalebox{.5}{MGD}}~,
\end{equation}
where $\tilde{\mathcal{L}}^{\scalebox{.5}{MGD}} = 2\pi\, g(\uprho)\left[4h(\uprho)(1-y^2)
\dot{\uprho}^2+g(\uprho)^2\right]^{1/2}$ and, once again, $y=\cos\theta$ is employed, in such a way that $\uprho\equiv\uprho(y)$. The minimization of Eq.~\eqref{area-min} with  respect to $y$, namely, $\delta A=0$, gives the following  ODE:
\begin{equation}\label{deMGD-almost}
    2(y^2-1)fg\ddot{\uprho}+8y(1-y^2)f^3\dot{\uprho}^3 + (1-y^2)\left[5f\frac{\mathrm{d}g}{\mathrm{d}\uprho}
    -g\frac{\mathrm{d}f}{\mathrm{d}\uprho}\right]\dot{\uprho}^2 + 4yfg\dot{\uprho}+g\frac{\mathrm{d}g}{\mathrm{d}\uprho}=0~,
\end{equation}
where the notation $g= g(\uprho)$ and $f= f(\uprho)$ was employed for simplicity. 
To solve Eq. \eqref{deMGD-almost}, the perturbative method must be applied, due to the lack of an analytical solution. For this purpose, the following expansion is then adopted, 
\begin{equation}\label{rho-almost}
    \uprho(y)=\varepsilon\uprho_1(y)+\varepsilon^2\uprho_2(y)~,
\end{equation}
with $\uprho_1(y_0)=\sqrt{\mathring{r}}$ and $\uprho_2(y_0)=0$, with boundary condition $\uprho(y_0)=0$.

The  $0^{\rm th}$-order term in Eq.~\eqref{rho-almost} is absent to avoid an area that is greater than one, at the point $(\uprho_0, \theta_0)$. In Eq.~\eqref{area-min} the constraint $\uprho<\uprho_0$ defines a  consistent value of the area. 
Therefore, looking for the $\rho$-functions up to second order, we insert Eq.~\eqref{rho-almost} into Eq.~\eqref{deMGD-almost}. It yields, at $1^{\rm st}$-order in $\varepsilon$, the expression 
\begin{equation} \label{1deMGD}
(y^2-1)\ddot{\uprho}_1+2y\dot{\uprho}_1+(1+\alpha)\uprho_1=0~,
\end{equation}
where $\alpha\equiv4\ell/\mathring{r}$. The solution of  Eq.~\eqref{1deMGD} reads $
    \uprho_1(y)=\mathcal{C}_1\mathrm{P}_\eta(y)$, 
with $\mathcal{C}_1=\sqrt{\mathring{r}}/\mathrm{P}_\eta(y_0)$, $\eta=\frac12\left(-1+\sqrt{-(3+4\alpha)}\right)$,  and $\mathrm{P}_\eta(y)$ is a Legendre polynomial of first kind. Such solution presents regularity at $y=1$ and has boundary condition $\uprho_1(y_0)=\sqrt{\mathring{r}}$~.

At $2^{\rm nd}$-order in $\varepsilon$,  Eq.~\eqref{deMGD-almost} is then a Legendre equation similar to Eq. (\ref{1deMGD}), 
\begin{equation}\label{2ordeMGD}
(y^2-1)\ddot{\uprho}_2+2y\dot{\uprho}_2+(1+\alpha)\uprho_2=0~,
\end{equation}
with $\uprho_2(y)=\mathcal{C}_2\mathrm{P}_\eta(y)$. Notwithstanding, the boundary condition $\uprho_2(y_0)=0$ demands $\mathcal{C}_2=0$. Thus $\uprho_2(y)=0$, leaving only  the $1^{\rm st}$-order in $\varepsilon$. 

\begin{figure}[htbp!]
    \centering
    \includegraphics[scale=.5]{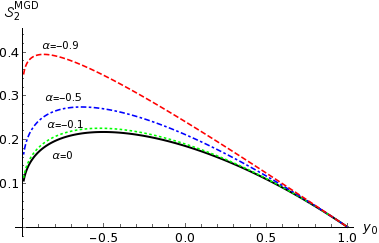}
    \caption{The evolution of the HEE $2^{\rm nd}$-order correction, in units of $\pi\mathring{r}\uprho_0^2$, related to the MGD parameter, according to the size of the subsystem.}
    \label{fig:S2MGDalmost}
\end{figure}

With the $\uprho$-functions, we can compute and analyze the area of the entangling surface. First, the expansion of the integrand in Eq.~\eqref{area-min} is adopted after the appropriate expansion in $\varepsilon$,  
\begin{equation}\label{ert}
\tilde{\mathcal{L}}^{\scalebox{.5}{MGD}}=2\pi \mathring{r}^2 +4\pi \mathring{r}\left[(1-y^2)\dot{\uprho}_1^2+\uprho_1^2\right]\varepsilon^2+\cdots.
\end{equation}
Inserting Eq. (\ref{ert}) into Eq.~\eqref{area-min} and executing the expansion of ${A}$, which reads $
{A}=A_0+A_1+A_2+\ldots~$, that is, the expansion of $\tilde{\mathcal{L}}^{\scalebox{.5}{MGD}}$, implying that corresponding HEE corrections yield
\begin{eqnarray}
\mathcal{S}_0^{\scalebox{.5}{MGD}} &=& \frac{\pi\mathring{r}^2}{2}\left(1-y_0\right),\nonumber\\
\mathcal{S}_1^{\scalebox{.5}{MGD}}&=& 0,\nonumber\\
\label{S2MGDalmost}
\mathcal{S}_2^{\scalebox{.5}{MGD}}&=&\frac{\pi \mathring{r}\uprho_0^2}{\mathrm{P}_\eta^2(y_0)}
\int_{y_0}^1 \mathrm{d}y\,\left[\left(\frac{1-y^2}{1+\alpha}\right)\dot{\mathrm{P}}_\eta^2(y)+\mathrm{P}_\eta^2(y)\right]. 
\end{eqnarray}
The calculation of $\mathcal{S}_2^{\scalebox{.5}{MGD}}$ is awkward enough to handle analytically. For solving it numerically, we plot the $\mathcal{S}_2^{\scalebox{.5}{MGD}}$ function in Fig.~\ref{fig:S2MGDalmost}, for different values of $\alpha$.

With the MGD parameter $\ell=0$, meaning $\alpha=0$, one recovers the HEE $2^{\rm nd}$-order correction for a Schwarzschild black hole. As the MGD parameter $\ell$ increases, one can observe the displacement -- upwards and to the left -- of the maximum of this order of correction looking at Fig.~\ref{fig:S2MGDalmost}, as $y_0$ decreases. This means that the MGD HEE $2^{\rm nd}$-order correction increases simultaneously to the requirement of the extension of the range of integration, that is, the size of the dual quantum subsystem.

\section{HEE in EMGD spacetimes}\label{sect:EMGD}

\quad\; As the HEE was already scrutinized in the last section for the MGD solution, the next step is to analyze the HEE for the EMGD metrics, where the notations  EMGD$_1$ and EMGD$_2$ are adopted for the $k=1$ and $k=2$ cases, respectively.

\subsection{EMGD \texorpdfstring{$k=1$}{Lg} case}\label{subsect:EMGD1}

\quad\; The EMGD \texorpdfstring{$k=1$}{Lg} case, represented by the solution in Eq.~\eqref{eq:k=1}, deals with the ADM mass $\mathbb{M}_1$ and the tidal charge $\mathbb{Q}_1$, being a Reissner-Nordstr\"om-like  metric. 

\subsubsection{Far from the horizon}\label{subsubsect:EMGD1beyond}

\quad\; Considering such boundaries far away from the horizon, the outcomes for the HEE corrections are similar to those ones found in Ref. \cite{Sun:2016dch}, once the direct replacements $M\mapsto\mathbb{M}_1$ and $Q^2\mapsto\mathbb{Q}_1$ {-- up to the $2^{\rm nd}$-order correction of HEE--} emulate the results presented in \cite{Sun:2016dch}. Therefore, the $1^{\rm st}$ and $2^{\rm nd}$-order corrections read
\begin{align}
\!\!S_1^{\scalebox{0.4}{EMGD$_1$}} &=\frac{\pi}{2}\mathbb{M}_1\left(1-y_0\right)^2 r_\infty\label{eq:S1EMGD}~,\\
\!\!\!S_2^{\scalebox{0.4}{EMGD$_1$}} &= \frac{\pi}{8} \left\{\mathbb{M}_1^2 \left[(7-y_0)(y_0-1)+2\log (y_0)+16\log\left(\frac{2}{1+y_0}\right)\right]+\mathbb{Q}_1\left[1-y_0^2 +2\log(y_0)\right]\right\}\label{eq:S2EMGD}~.
\end{align}
We opt not to display the $0^{\rm th}$-order, as it is the same as the one presented in Ref. in \cite{Sun:2016dch}, being  independent of the ADM mass $\mathbb{M}_1$, for this case. 

\begin{figure}[htbp!]
\begin{center}
\includegraphics[scale=0.5]{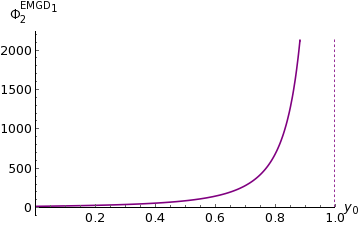}
\end{center}
\caption{Global profile of the factor between the HEE $2^{\rm nd}$-order corrections with respect to $y_0$.} \label{fig:phi2EMGD1}
\end{figure}

Assigning the ADM mass $\mathbb{M}_1$ and tidal charge $\mathbb{Q}_1$ to the mass parameter $M$, which is the black hole Misner--Sharp mass function in the Reissner-Nordstr\"om metric, the contribution from the MGD can be then closer investigated. Hence, after those respective identifications, one gets  
\begin{equation}
S_2^{\scalebox{.5}{EMGD$_1$}} = \pi M^2 \left[-y_0^2+4y_0-3+2\log(y_0)+8\log\left(\frac{2}{1+y_0}\right)\right]~.
\end{equation}
Thus, the corrections to the  HEE can be compared to the Schwarzschild solution. For this task, in compliance with what has been established in Sect.~\ref{sect:MGD-setup}, that is, $\mathbb{M}_1=2M$ and $\mathbb{Q}_1=4M^2$, we determine the following factors between each order of correction to the HEE. First, Eq.~\eqref{Phi_n} yields $\Phi_0^{\scalebox{.5}{EMGD$_1$}}=1$ and $\Phi_1^{\scalebox{.5}{EMGD$_1$}}=2$. Hence, the $2^{\rm nd}$-order corrections may be written as
\begin{equation}
\Phi_2^{\scalebox{.5}{EMGD$_1$}}=8\left[\frac{y_0^2-4y_0+3-2\log (y_0)-8\log\left(\frac{2}{1+y_0}\right)}{y_0^2-8y_0+7-2\log(y_0)-16\log\left(\frac{2}{1+y_0}\right)}\right]~.\label{eq:phi2EMGD1}
\end{equation}
Such factor varies independently of the mass parameter, $M$, and $\lim_{y_0\to0} \Phi_2^{\scalebox{.5}{EMGD$_1$}}=8$, whereas $\lim_{y_0\to1}\Phi_2^{\scalebox{.5}{EMGD$_1$}} \to\infty$.  Fig. \ref{fig:phi2EMGD1} shows the global profile of this factor.

\begin{figure}[htbp!]
\begin{center}
\includegraphics[scale=0.5]{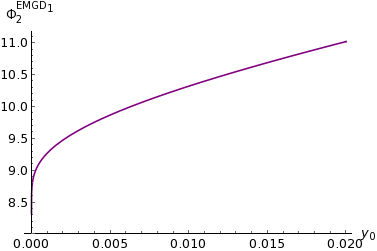}
\end{center}
\caption{Profile of the factor between the HEE $2^{\rm nd}$-order corrections when $y_0$ is close to $0$.} \label{fig:phi2EMGD1_closeto0}
\end{figure}

The $\Phi_2^{\scalebox{.5}{EMGD$_1$}}$ function is not monotonic, presenting an inflection point. Looking closely to values of $y_0$,  one may observe the transitions from an initial increment to an intermediate lowering, and next, increases again. Fig. \ref{fig:phi2EMGD1_closeto0}, which magnifies Fig. \ref{fig:phi2EMGD1} for $y_0$ near the origin, displays this behavior.
In addition, there are two brief and important features to emphasize.  Firstly,  at the first sight, inspecting Eq.~\eqref{eq:S2EMGD} and setting $\mathbb{Q}_1\to 0$, one promptly verifies that the $2^{\rm nd}$-order correction, considering the ADM mass related to the mass parameter $M$, is four times the same order correction to the Schwarzschild spacetime. Second, the HEE $2^{\rm nd}$-order correction in the EMGD$_1$ case, related to the mass parameter $M$, is always negative. It can be interpreted as an increment of attenuation in the entropy function, as the HEE $2^{\rm nd}$-order correction in the  Schwarzschild spacetimes is also negative.

\begin{figure}[htbp!]
\begin{center}
\includegraphics[scale=0.5]{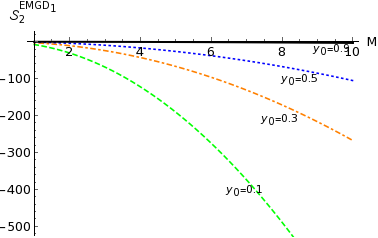}
\end{center}
\caption{Behavior of the HEE $2^{\rm nd}$-order correction in EMGD$_1$ spacetime related to the mass parameter $M$ corresponding to different values of $y_0$.} \label{fig:S2EMGD1}
\end{figure}

Hereon, let us take a look at the mass parameter after choosing a specific size of the entangling surface, which means to delimitate the minimal area. For $y_0$ values close to zero, the increment of the mass parameter $M$, accentuates the $2^{\rm nd}$-order contribution for  EMGD$_1$, when one  works with an entangling surface with a specific size. On the  other hand, there is no such accentuation when the $y_0$ integration limit equals 1, even when the black hole mass increases. It is very illustrative to display the profile of such correction in Fig.~\ref{fig:S2EMGD1}, to compare with the same order of correction of the Schwarzschild black hole displayed in Fig.~\ref{fig:S2Schwarz}.
As the black hole mass increases, the attenuation becomes greater. In addition, the attenuation increases faster for small values of $y_0$. Otherwise, the attenuation continues to increase in a slower rate.
Let us implement the same procedure for the $2^{\rm nd}$-order correction in EMGD$_1$ related only to $M$.

\begin{figure}[htbp!]
\begin{center}
\includegraphics[scale=0.5]{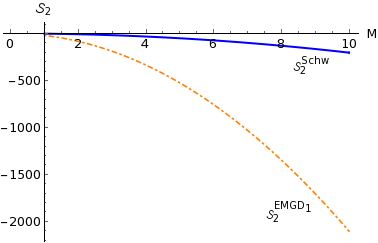}
\includegraphics[scale=0.5]{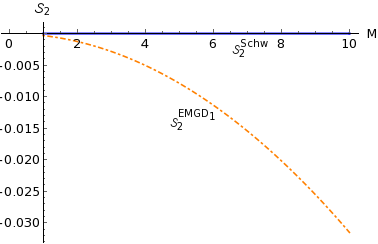}
\caption{The difference between the HEE corrections for large [small] entangling surfaces, initiating at $y_0=0.01$ [$y_0=0.99$].}\label{fig:S2Schwarz_EMGD1_great_and_small_y}
\end{center}
\end{figure}

One can notice that the same analysis can be accomplished to the EMGD$_1$ related only to $M$. Moreover, the attenuation is more intense in the EMGD$_1$ case, when compared to the Schwarzschild one. It is also worth to emphasize that such analysis considered the tidal charge and the ADM mass as functions of the mass parameter $M$. To clarify this point, we take two values for $y_0$, one of them close to 0 and another one close to 1, displaying both corrections in  Fig.~\ref{fig:S2Schwarz_EMGD1_great_and_small_y}. 

\begin{figure}[htbp!]
\begin{center}
\includegraphics[scale=0.5]{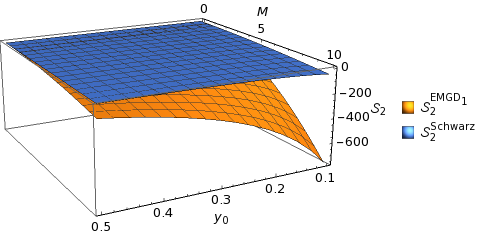}
\end{center}
\caption{The HEE corrections for different values of $M$ and $0<y_0<0.5$.} \label{fig:S2EMGD1_M_and_y0}
\end{figure}

Finally, both corrections can be plotted making $M$ and $y_0$ to run in their specific ranges, as shown in Fig. \ref{fig:S2EMGD1_M_and_y0}. As one can observe, a more restrictive interval for $y_ 0$ is considered, to realize the profile of each minimal surface.

\noindent It is straightforward to observe how the range of integration characterized by $y_0$ establishes a major difference between both $2^{\rm nd}$-order corrections, as the black hole mass increases. On  the other hand, the difference is insignificant when the size of the minimal surface is reduced as $y_0$ increases.

\subsubsection{Almost on the horizon}\label{subsubsect:EMGD1almost}

From now on, we initiate the analysis of the EMGD$_1$ black hole entropy, concerning the boundary almost on the horizon. The solution for this case is based on the metric in Eq.~\eqref{eq:k=1}. According to Ref. \cite{Casadio:2015gea}, this metric corresponds to an extremal black hole, which has degenerate horizons represented by $\mathring{r}=\mathbb{M}_1$. In this sense, the functions
\begin{equation}
    \mathrm{e}^\nu=\mathrm{e}^{-\lambda}=    \frac{\left(r-\mathring{r}\right)^2}{r^2}~,
\end{equation}
describe the constant $t$-fold induced metric as
\begin{equation}\label{eq:ind_metric1}
    \mathrm{d}s^2=\mathrm{p}(\uprho)\mathrm{d}\uprho^2+\mathrm{q}(\uprho)^2(\mathrm{d}\theta^2+\sin^2\theta \mathrm{d}\varphi^2)~,
\end{equation}
which is built with the variable change $\mathrm{q}(\uprho)=\uprho^2+\mathring{r}$. Above, one also denotes $\mathrm{p}(\uprho)=\frac{4(\uprho^2+\mathring{r})^2}{\uprho^2}$. Proceeding to the computation of the area functional and its minimization yields the highly nonlinear ODE, 
\begin{equation}\label{eq:full-edo-emgd1-almost}
    \dot{\uprho}(1-y^2)\left[-2\mathrm{p}\mathrm{q}^2\ddot{\uprho}+2y\mathrm{p}^2\dot{\uprho}^3
-\left(\dot{\mathrm{p}}\mathrm{q}^2-6\mathrm{p}\mathrm{q}\dot{\mathrm{q}}\right)
\dot{\uprho}\right]
 +4y\mathrm{p}\mathrm{q}^2 \dot{\uprho}^2 + 4\mathrm{q}^3\dot{\mathrm{q}}=0.
\end{equation}

\begin{figure}[htbp!]
    \centering
    \includegraphics[scale=0.5]{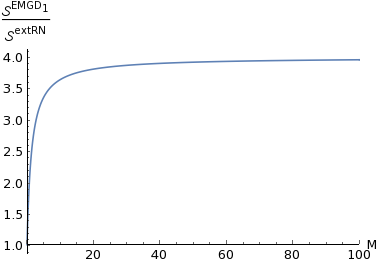}
    \caption{The ratio of the HEE in EMGD$_1$ spacetime to HEE in the extremal RN one. $M$ has units of $\uprho_0^2$ and $M/\uprho_0^2\in\left[10^{-2},10^2\right]$.}
    \label{fig:ratioSEMGD1almost}
\end{figure}

Next, similar steps implemented from Eq.~\eqref{area-min} to  Eq.~\eqref{rho-almost} will be employed. In fact, it consists of a perturbation procedure to obtain an approximated solution up to $2^{\rm nd}$-order of Eq.~\eqref{eq:full-edo-emgd1-almost}. The expanded ODE  is awkward and difficult to solve through analytical methods. On the other hand, one can look at the  $0^{\rm th}$-order in $\varepsilon$, which is
\begin{equation}\label{eq:0th-edo-emgd1-almost}
\left(1-y^2\right)\left(-\uprho_1^2 \ddot{\uprho}_1 + 4y\dot{\uprho}_1^3 + \uprho_1 \dot{\uprho}_1^2\right)+2 y \uprho_1^2\dot{\uprho}_1 =0~.
\end{equation}
We employ the boundary conditions, constraining Eq.~\eqref{rho-almost},  to filter the infinite possible analytical solutions to Eq.~\eqref{eq:0th-edo-emgd1-almost}, implying that 
\begin{equation}\label{eq:k=1-almost-rho}
    \uprho(y)=\uprho_0.
\end{equation}
In full agreement with \cite{Sun:2016dch}, such constant solution is the only one that attends strictly the boundary condition. It disposes quite differently of the Schwarzschild or MGD spacetimes looking for a minimal surface almost on the horizon. Such so restrictive solution only could emphasize that Eq.~\eqref{eq:full-edo-emgd1-almost} needs to be investigated at higher orders, once the constant solution shown by Eq.~\eqref{eq:k=1-almost-rho} is not a solution of the full Eq.~\eqref{eq:0th-edo-emgd1-almost}. Finally, we reinforce the solution Eq.~\eqref{eq:k=1-almost-rho} as a completely safe one, up to $2^{\rm nd}$-order. Thus, with the solution \eqref{eq:k=1-almost-rho}, we are able to estimate the entropy as follows:
\begin{equation}\label{eq:k=1-area-almost}
\mathcal{S}^{\scalebox{.5}{EMGD$_1$}}=\frac{\pi}{2}\int_{y_0}^1\mathrm{d}y \mathrm{q}\left[(1-y^2)\mathrm{p}\dot{\uprho}^2+\mathrm{q}^2\right]^\frac{1}{2}
\end{equation}
and  Eq.~\eqref{eq:k=1-almost-rho} yields 
\begin{equation}\label{eq:SEMGD1almost}
    \mathcal{S}^{\scalebox{.5}{EMGD$_1$}}=\frac{\pi}{2}\left(1-y_0\right)\left(\uprho_0^2+\mathring{r}\right)^2\equiv \frac{\pi}{2}\left(1-y_0\right)\left(\textsc{R}_{\scalebox{.5}{Bound}}^{\scalebox{.5}{EMGD$_1$}}\right)^2~,
\end{equation}
which has $\textsc{R}_{\scalebox{.5}{Bound}}^{\scalebox{.5}{EMGD$_1$}}=\uprho_0^2+\mathring{r}$ representing the boundary surface radius. Since $\mathring{r}=\mathbb{M}_1=2M$, the entropy is increased, compared to that one established for the extremal RN black hole in Ref. \cite{Sun:2016dch}. Such an entropy increment  is explicit through the ratio
\begin{equation}\label{eq:ratioSEMGD1toRN}
 \frac{\mathcal{S}^{\scalebox{.5}{EMGD$_1$}}}{\mathcal{S}^{\scalebox{.5}{extRN}}}=\left(\frac{\textsc{R}_{\scalebox{.5}{Bound}}^{\scalebox{.5}{EMGD$_1$}}}{\textsc{R}_{\scalebox{.5}{Bound}}^{\scalebox{.5}{extRN}}}\right)^2= \left(\frac{1+\frac{2M}{\uprho_0^2}}{1+\frac{M}{\uprho_0^2}}\right)^2 ~,
\end{equation}
standing $\mathcal{S}^{\scalebox{.5}{extRN}}=\pi/2\left(1-y_0\right)\left(\uprho_0^2+M\right)^2$ as the entropy of an extremal RN black hole, where the horizon is $\mathring{r}_{\scalebox{.5}{extRN}}=M$~. With that, we obtain the entropy gain without any mention to the range of the minimal surface. Importantly, the ratio is positive, indicating the increment of the entropy in the EMGD$_1$ scenario for extremal black holes. Fig.~\ref{fig:ratioSEMGD1almost} points out such profile.

Fixing $\uprho_0^2$ provides a first range with a fast-growing entropy until $M=10\uprho_0^2$. After this, there is a very slow-growing, stabilizing at a ratio equal to $4$. On the one hand, it does not matter how large the black hole is, the ratio stabilizes at $4$, even with the displacement of the extremal horizon in the EMGD$_1$ case. On the other hand, entropies of black holes with $10^{-2}\uprho^2_0\lesssim M\lesssim 20\uprho_0^2$ have meaningful increments, which shows simply and directly the contribution from the EMGD$_1$ approach.

\subsection{EMGD \texorpdfstring{$k=2$}{Lg} case}\label{subsect:EMGD2}

\quad\;We settle here an analogue construction to the one in Sect.~\ref{subsect:MGDbeyond}, using the EMGD metric with the temporal and radial components respectively given in Eqs.~\eqref{eq:k=2nu} and \eqref{eq:k=2lambda-minus}, for $k=2$. 

\subsubsection{Far from the horizon}\label{subsubsect:EMGD2beyond}

\quad\; Let us consider the steps in Eqs.~\eqref{eq:area} and \eqref{eq:Lag1}. The replacement $\mathcal{L}^{\scalebox{.5}{MGD}}\mapsto\mathcal{L}^{\scalebox{.5}{EMGD$_2$}}$ is then necessary, as a distinct $\mathcal{F}$ must be taken into account. In fact, using the EMGD$_2$ metric, one gets a quite similar ODE shown in Eq.~\eqref{eq:EOM}, which is the metric radial component. Once again, that similar Eq.~\eqref{eq:EOM} with the current $\mathcal{F}$ must be solved perturbatively. Before it, one establishes the parameter of expansion $\varepsilon=\mathbb{M}_2/r_\infty$ and the corresponding parameters $\mathbb{Q}_2=\kappa^2\mathbb{M}_2^2$ and $\textsc{s}=\omega\mathbb{M}_2$~, that  follow a similar reasoning of the previous cases\footnote{The only difference here is the use of $\kappa^2$ instead of $\kappa$, as the auxiliary parameter in the corresponding expansion parameter $\mathbb{Q}_2$. Since there is no numerical difference, we adopt this form to follow the same exponentiation of the mass term $\mathbb{M}_2$.}.

Applying a similar procedure realized in Sect.~\ref{subsect:MGDbeyond}, we need to compute the auxiliary $g$-functions for the series expansions necessary to find the $r$-functions,  which are crucial to calculate the HEE corrections up to $2^{\rm nd}$-order. Hence, applying the expansions (\ref{eq:Fexp}, \ref{eq:rexp}), we find the $g$-functions:
\begin{align}
    g_1(y) & = \frac{y r_\infty}{3w_0}\left(3\omega -2\right)~,\\
    g_2(y) & = \frac{y^2 r_\infty}{18 w_0^2} \left[\left(15 \kappa -18 \omega ^2+15 \omega -16\right) r_\infty+6(2-3 \omega ) r_1(y)\right]~,
\end{align}
that are necessary functions to find the respective ODEs that lead us to determine the $r_1(y)$ and $r_2(y)$ functions. Each one of them is solved strictly as engaged in Sect. \ref{subsect:MGDbeyond}, using the boundary conditions to compute the constants of integration for each function. Hence, at $1^{\rm st}$ and $2^{\rm nd}$ orders, as follows, it implies  respectively that 
\begin{equation}
  \ddot{r}_1(y) + \frac{\left(5 y^2-3\right)}{y \left(y^2-1\right)}\dot{r}_1(y)+\frac{\left(3 y^2-1\right)}{y^2\left(y^2-1\right)}r_1(y)  = \frac{\left(3 y^2+1\right)\left(3\omega -2\right)r_\infty}{6y^2\left(1-y^2\right)}~,
\end{equation}
whose solution is
\begin{equation}
    r_1(y)=\frac{(2-3\omega)r_\infty}{y}\left[y-y_0-2\log\left(\frac{1+y}{1+y_0}\right) +2\log\left(\frac{y}{y_0}\right)\right]~;
\end{equation}
and
\begin{equation}
 \ddot{r}_2(y) + \frac{\left(5 y^2-3\right)}{y \left(y^2-1\right)}\dot{r}_2(y)+\frac{\left(3 y^2-1\right)}{y^2\left(y^2-1\right)}r_2(y) = \mathcal{R}(y)
\end{equation}
with
\begin{equation}
   \mathcal{R}(y) = \frac{r_\infty^2}{18w_0}\left[\frac{y^3 \left(30\kappa^2-9\omega^2-6\omega - 20\right) + \left(4-3y\right)\left(2-3\omega\right)^2}{ y^2 \left(1-y^2\right)}\right]~,
\end{equation}
which has  solution given by 
\begin{equation}
 r_2(y) = \frac{r_\infty^2}{144w_0 y}\left[\left(y^2-y_0^2\right)\textsc{V}_1(\kappa,\omega) - 2\textsc{V}_2(\kappa,\omega)\log\left(\frac{y}{y_0}\right) + \textsc{V}_3(\omega)\log\left(\frac{1+y}{1+y_0}\right)\right] ~.
\end{equation}
with $\textsc{V}_1(\kappa,\omega)=20-30\kappa^2+9\omega^2 + 26\omega$~,~$\textsc{V}_2(\kappa,\omega)=20+30\kappa^2 + 81\omega^2 - 126\omega$~,~ and $\textsc{V}_3(\omega)=32(3\omega-2)^2$~.

Once again, we use the $r$-functions to proceed with the expansion of $\mathcal{L}^{\scalebox{.5}{EMGD$_2$}}$ towards the computation of the area and, consequently, the HEE expression up to $2^{\rm nd}$-order. Thereupon, the $0^{\rm th}$- and $1^{\rm st}$-order of the HEE corrections are, respectively,
\begin{align}
 \mathcal{S}_0^{\scalebox{.5}{EMGD$_2$}} &= \frac{\pi w_0^2}{4}\left(\frac{1}{y_0^2}-1\right)\label{S0EMGD2beyond}~,\\
 \mathcal{S}_1^{\scalebox{.5}{EMGD$_2$}} &= \frac{\pi r_\infty\mathbb{M}_2}{4}\left(1-y_0\right)^2 \left(\frac{2}{3} - \frac{\textsc{s}}{\mathbb{M}_2}\right)\label{S1EMGD2beyond}~.
\end{align}
It is worth to emphasize that Eq.~\eqref{S1EMGD2beyond} has the presence of the EMGD$_2$ parameter. It is quite different,  compared with the $k=1$ case, where there is no EMGD$_2$ parameter in such order of correction. One can notice a growth like the $1^{\rm st}$-order correction from \cite{Sun:2016dch} as well as succeeded in the EMGD$_1$. This occurs due to the ADM mass, which corresponds to $3M$ in this $k=2$ case. Hence, $\mathcal{S}_1^{\scalebox{.5}{EMGD$_2$}}=\mathcal{S}_1^{\scalebox{.5}{MGD}}$. Again, there is no contribution from the charge as well as one noticed in Ref. \cite{Sun:2016dch} to RN spacetimes.

Carrying on, the $2^{\rm nd}$-order of the HEE correction reads 
\begin{eqnarray}
    \mathcal{S}_2^{\scalebox{.5}{EMGD$_2$}} &=& \frac{\pi}{288}\left\{4\mathbb{M}_2^2 \left[\left(y_0-1\right)\left(3 y_0+11\right) - 6\log (y_0)+16\log\left(\frac{2}{1+y_0}\right)\right]\right. \nonumber\\
     &+&\left. 30\mathbb{Q}_2\left[1 - y_0^2+2\log(y_0)\right]+ 9\textsc{s}^2 \left[\left(1-y_0\right)\left(y_0-7\right) + \log\left(\frac{65536\,y_0}{(1+y_0)^{16}}\right)\right]\right. \nonumber\\
    &+&\left.  6\textsc{s}\mathbb{M}_2\left[\left(y_0-1\right)\left(5 y_0-11\right) - 10\log(y_0) - 32\log \left(\frac{2}{y_0+1}\right)\right] \right\}~.\label{S2EMGD2beyond}
\end{eqnarray}

Looking at the previous cases, the MGD and EMGD$_1$, there is a leading difference here. Even in the $\textsc{s}\to0$ regime, there is a numerical difference, when compared to  the EMGD$_1$. It would be nice to plot some comparison with the Schwarzschild black hole or, strictly, with the RN without the $\textsc{s}$ parameter, to scale the numerical contribution.

\begin{figure}[htbp!]
    \centering
    \includegraphics[scale=0.5]{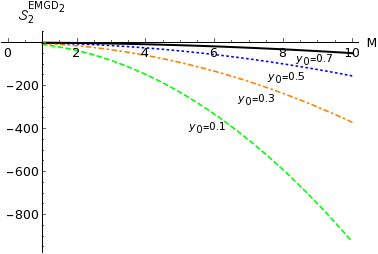}
    \includegraphics[scale=0.5]{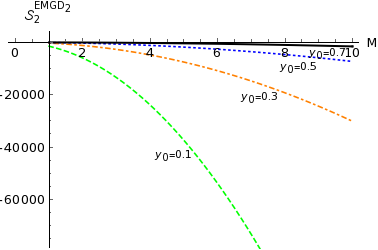}
    \caption{Behavior of the HEE $2^{\rm nd}$-order correction depending on $M$ for fixed values of $y_0$. We adopt $\zeta=-0.1$ (heavy tension) in the plot on the left, while $\zeta=-100$ (light tension) was adopted on the right.}
    \label{fig:S2EMGD2_M_zeta-01and-100}
\end{figure}

Following an analogue procedure established in EMGD$_1$ case, let us put the ADM mass and the tidal charge in terms of Schwarzschild mass parameter, which are $\mathbb{M}_2=3M$ and $\mathbb{Q}_2=12M^2$, respectively. Besides, we use $\zeta=\textsc{s}/M$ as well as it has been done in the MGD case. Over again, the main purpose here is also fixing $M$ to analyze the influence of a finite brane tension at this order of HEE correction. Continuing, the expression below carries only the lower-limit of the integration in the area functional and the parameter $\zeta$. In this sense, we clearly could investigate the ratio related with the $2^{\rm nd}$-order correction for Schwarzschild spacetimes, that is,
 \begin{equation}
    \tilde{\mathcal{S}}_2^{\scalebox{.5}{EMGD$_2$}} =\frac{\pi M^2}{32}\left[\textsc{W}_3\left(\zeta ,y_0\right)+ 2\textsc{W}_2\log\left(y_0\right) + 2\textsc{W}_3\log\left(\frac{2}{1+y_0}\right)\right]~,
 \end{equation}
where
\begin{align}
    \textsc{W}_3\left(\zeta,y_0\right) &= \textsc{W}_1(\zeta) y_0 - \textsc{W}_2(\zeta) y_0^2 + \left(22-7\zeta\right)\zeta-4,\\
    \textsc{W}_2(\zeta) &= 28-10\zeta+\zeta^2, \\
    \textsc{W}_3(\zeta) &= 8\left(\zeta-2\right)^2~.
\end{align}
Next, Fig.~\ref{fig:S2EMGD2_M_zeta-01and-100} illustrates, for two values of $\zeta$ -- the first one representing a high brane tension and another one depicting a low brane tension -- how the size of the minimal\footnote{The importance of the range in the integration to obtain the entropy through the area functional can be analyzed as follows. According to Ref. \cite{Sun:2016dch}, HEE is a short form to calculate the entanglement entropy of a subsystem in the dual theory. Therefore, $y_0$ defines uniquely the size of the subsystem.} surface affects the $2^{\rm nd}$-order correction.

\begin{figure}[htbp!]
    \centering
    \includegraphics[scale=0.5]{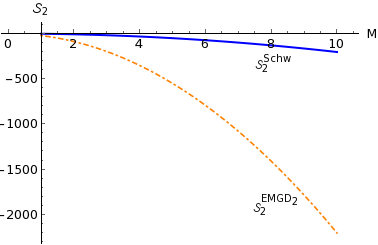}
    \includegraphics[scale=0.5]{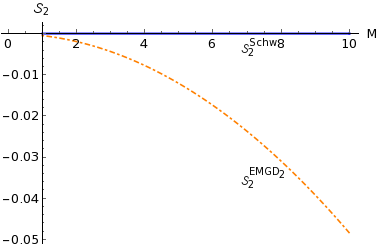}
    \caption{Behavior of the HEE $2^{\rm nd}$-order correction depending on $M$ for $\zeta=-0.1$ and setting $y_0=0.01$ [$y_0=0.99$] on the left [right].}
    \label{fig:S2EMGD2_S2Schw_M_and_zeta-01_y0_closeto0and1}
\end{figure}

\begin{figure}[htbp!]
    \centering
    \includegraphics[scale=0.5]{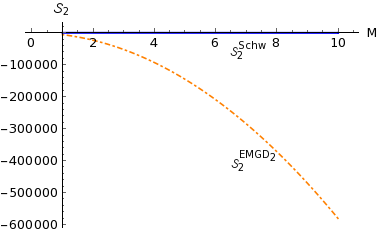}
    \includegraphics[scale=0.5]{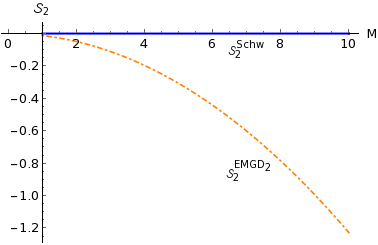}
    \caption{Behavior of the HEE $2^{\rm nd}$-order correction depending on $M$ for $\zeta=-100$ and setting $y_0=0.01$ [$y_0=0.99$] on the left [right].}
    \label{fig:S2EMGD2_S2Schw_M_and_zeta-100_y0_closeto0and1}
\end{figure}

\begin{figure}[htbp!]
    \centering
    \includegraphics[scale=0.5]{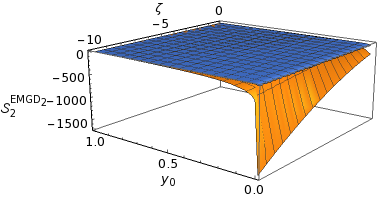}
    \caption{The $2^{\rm nd}$-order corrections to EMGD$_2$ and Schwarzschild spacetimes pondering light and heavy tension on the brane as well as the full range of $y_0$.}
    \label{S2EMGD2_zeta_y0}
\end{figure}

\begin{figure}[htbp!]
    \centering
    \includegraphics[scale=0.5]{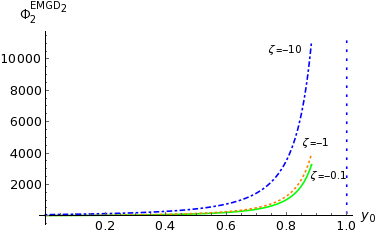}
    \caption{The $2^{\rm nd}$-order correction of the HEE in EMGD$_2$ for fixed values of $\zeta$.}
    \label{fig:phi2EMGD2}
\end{figure}

To compare the HEE $2^{\rm nd}$-order correction in EMGD$_2$ to the one of a Schwarzschild one, for different $M$ values, we set $\zeta=-0.1$, in Fig.~\ref{fig:S2EMGD2_S2Schw_M_and_zeta-01_y0_closeto0and1}, and $\zeta=-100$ in Fig.~\ref{fig:S2EMGD2_S2Schw_M_and_zeta-100_y0_closeto0and1}, specifying two kinds of ranges: a first one close to 0, and another one close to 1.
\newpage
To analyze a wide range scenario to $\zeta$ and $y_0$, we plot  Fig.~\ref{S2EMGD2_zeta_y0}.
Besides, the ratio to this order is
\begin{equation}\label{phi2EMGD2}
    \Phi_2^{\scalebox{.5}{EMGD$_2$}} = \frac{1}{4}\left\{28-10\zeta+\zeta^2-48\left(\zeta-4\right)\left[\frac{y_0 - 1 + 2\log\left(\frac{2}{1+y_0}\right)}{-7+8y_0-y_0^2 \log\left(\frac{65536\,y_0}{(1+y_0)^{16}}\right)}\right]\right\}~,
\end{equation}
which graphically is presented in Fig.~\ref{fig:phi2EMGD2}.

\begin{figure}[htbp!]
    \centering
    \includegraphics[scale=0.5]{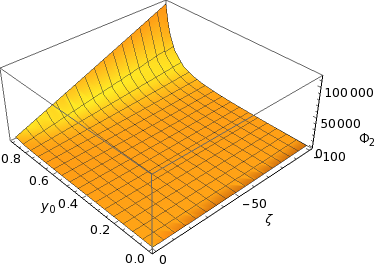}
    \caption{The $2^{\rm nd}$-order correction of the HEE in EMGD$_2$ for values of $\zeta$ and $y_0$.}
    \label{fig:phi2EMGD2_zeta_y0}
\end{figure}

In a general framework, leaving $\zeta$ and $y_0$ free to run within their valid interval of values, Fig.~\ref{fig:phi2EMGD2_zeta_y0} shows the $2^{\rm nd}$-order ratio.

For completeness, we establish
\begin{align}
    \Phi_0^{\scalebox{.5}{EMGD$_2$}} &=1~,\\
    \Phi_1^{\scalebox{.5}{EMGD$_2$}} &=1-\frac{\zeta}{2}~.
\end{align}
Note that both ratios above are identical to those ones obtained in the MGD case.

Some features can be extracted out of Eq.~\eqref{phi2EMGD2} and Fig.~\ref{fig:phi2EMGD2_zeta_y0}: (i) when the size of the minimal area is reduced, which is implemented with $y_0\gtrsim0.9$, a low brane tension hugely contributes to the increment of the ratio; (ii) when $y_0\to0$, the parameter related to the brane tension is dominant.

\subsubsection{Almost on the horizon}\label{subsubsect:EMGD2almost}

\quad\; Specifically, we now deal with the metric \eqref{eq:metric_general_spher_symm}, which carries the time component \eqref{eq:k=2nu} and the radial one \eqref{eq:k=2lambda-minus}, with coefficients $c_m$'s displayed in \eqref{k=2Coeffs}, as
\begin{equation}
    \textrm{e}^{-\lambda}=\left(\frac{1}{r-\mu\mathring{r}}\right)\sum_{m=0}^8\frac{c_m}{r^{m-1}}~,
\end{equation}
where $\mathring{r}=r_e=1.12\mathbb{M}_2$ stands for the degenerate event horizon determined in \cite{Casadio:2015gea} and $\mu\approx 0.4533$. We must implement the subtle displacement of the event horizon, that is, $r=\uprho+\mathring{r}$, $\uprho>0$, and fix the boundary on the horizon with $\uprho_0=\varepsilon\mathring{r}$ with $\varepsilon\ll 1$. Once again, the $\theta=\theta_0$ circumference maps the entangling surface. Hence, the resulting induced metric on the $t$-constant manifold is
\begin{equation}\label{eq:indMetric_k=2Almost}
\mathrm{d}\hat{s}^2=\left[\mathfrak{p}(\uprho)\left(
\frac{\mathrm{d}\uprho}{\mathrm{d}\theta}\right)^2+\mathfrak{q}(\uprho)^2\right]\mathrm{d}\theta^2
+\left[\mathfrak{q}(\uprho)\sin\theta\right]^2\mathrm{d}\varphi^2,
\end{equation}
where $r\mapsto\mathfrak{q}(\rho)$ and
\begin{equation}
 \mathfrak{p}(\uprho)=\left(\uprho+\mu\mathring{r}\right)\left(\uprho+\mathring{r}\right)^7\left[\sum_{m=0}^{8}c_m\left(\uprho+\mathring{r}\right)^{8-m}\right]^{-1}~.
\end{equation}

Finding $\uprho\equiv\uprho(\theta)$ means to minimize the surface area
\begin{equation}\label{area-minEMGD2almost}
A=\int_{y_0}^1 \mathrm{d}y ~\tilde{\mathcal{L}}^{\scalebox{.5}{EMGD$_2$}},
\end{equation}
where $\tilde{\mathcal{L}}^{\scalebox{.5}{EMGD$_2$}} = 2\pi \mathfrak{q}(\uprho)\left[\mathfrak{p}(\uprho)(1-y^2)
\dot{\uprho}^2+\mathfrak{q}(\uprho)^2\right]^{1/2}$, $y=\cos\theta$ is employed to attain $\uprho\equiv\uprho(y)$. The variation of Eq.~\eqref{area-min} with  respect to $y$, and taking $\delta A=0$, gives
\begin{equation}\label{deEMGD2almost}
    4\mathfrak{q}^3\frac{\mathrm{d}\mathfrak{q}}{\mathrm{d}\uprho} + 4y\mathfrak{p}\mathfrak{q}^2\dot{\uprho}+ (1-y^2)\left[\left(-\mathfrak{q}^2\frac{\mathrm{d}\mathfrak{p}}{\mathrm{d}\uprho}
    +6\mathfrak{p}\mathfrak{q}\frac{\mathrm{d}\mathfrak{q}}{\mathrm{d}\uprho}\right)\dot{\uprho}^2 + 2y\mathfrak{p}^2\dot{\uprho}^3 -2\mathfrak{p}\mathfrak{q}^2\ddot{\uprho}\right]=0~,
\end{equation}
where $\mathfrak{q}=\mathfrak{q}(\uprho)$, $\mathfrak{p}= \mathfrak{q}(\uprho)$. Now, the perturbation procedure previously used in Sect.~\ref{subsect:MGDalmost} is also applied here to build two ODEs  up to $2^{\rm nd}$-order in $\varepsilon$ by the expansion $\uprho=\varepsilon\uprho_1+\varepsilon^2\uprho_2$ into the Eq.~\eqref{deEMGD2almost}.  Such $\rho$-functions are important to execute the series expansion of the integrand in Eq.~\eqref{area-minEMGD2almost} up to $2^{\rm nd}$-order, which will be substantial to determine the HEE corrections in the present case. Thus, the first one of them, that is, the $1^{\rm st}$-order in $\varepsilon$ ODE is
\begin{equation}\label{EMGD2almostDE1}
    \left(y^2-1\right)\ddot{\uprho}_1+2y\dot{\uprho}_1+\gamma\uprho_1=0~,
\end{equation}
where
\begin{equation}\label{gammaParameter}
    \gamma=\frac{6}{\mu}\sum_{m=0}^{8}\frac{c_m}{\mathring{r}^m}~.
\end{equation}
Eq.~\eqref{EMGD2almostDE1} has the general solution
\begin{equation}
    \uprho_1(y)=\textsc{A}_1\mathrm{P}_\eta(y)+\textsc{A}_2\mathrm{Q}_\eta(y),
\end{equation}
with $\mathrm{P}_\eta(y)$ and $\mathrm{Q}_\eta(y)$ as Legendre polynomials of the first and second kind, respectively, and $\eta=1/2\left(-1+\sqrt{1-4\gamma}\right)$. Requiring regularity at $y=\pm1$, one needs to set $\textsc{A}_2=0$ since $\mathrm{Q}_\eta(y)$ is not regular in such points. The boundary condition $\uprho_0=\varepsilon\uprho_1(y_0)$ determines $\textsc{A}_1$ and leaves us with
\begin{equation}\label{solDE1EMGD2almost}
    \uprho_1\equiv\uprho_1(y)=\frac{\mathring{r}\mathrm{P}_\eta(y)}{\mathrm{P}_\eta(y_0)}~.
\end{equation}

The $2^{\rm nd}$-order ODE reads 
\begin{equation}\label{EMGD2almostDE2}
    \left(y^2-1\right)\ddot{\uprho}_2+2y\dot{\uprho}_2+\gamma\uprho_2 + \Omega(y,\gamma,\beta)=0~,
\end{equation}
where
\begin{equation}\label{HfunctionEMGD2almost}
    \Omega(y,\gamma,\beta)=\frac{1}{\mathring{r}}\left[\frac{1}{2}\left(\upbeta-8\right) \left(y^2-1\right)\dot{\uprho}_1^2+ \upbeta\left(y^2-1\right)\uprho_1\ddot{\uprho}_1 + \gamma\uprho_1^2 + 2\upbeta y\uprho_1\dot{\uprho}_1\right]
\end{equation}
and
\begin{equation}\label{betaParameter}
    \upbeta=9+\frac{1}{\mu}-\frac{\sum_{j=0}^{7}(8-j)c_j\mathring{r}^{(8-j)}}{\sum_{i=0}^{8}c_i\mathring{r}^{(8-i)}}~.
\end{equation}
Eq.~\eqref{EMGD2almostDE2} is a linear non-homogeneous ODE. The presence of the $\Omega(y,\gamma,\beta)$ permits a variety of solutions conditioned to the parameters $\upbeta$ and $\gamma$, which by themselves are constrained to the physical parameters of EMGD$_2$ case, \emph{i.e.}, the ADM mass $\mathbb{M}_2$, the tidal charge $\mathbb{Q}_2$ and the EMGD$_2$ parameter $\textsc{s}$ within $c$-coefficients explicitly detailed in \eqref{k=2Coeffs}. Therefore, the general analytical solution for Eq.~\eqref{EMGD2almostDE2} is written as
\begin{equation}\label{eq:rho2EMGD2almost}
    \uprho_2(y)=\textsc{B}_1\textsc{P}_\eta(y) + \textsc{B}_2\textsc{Q}_\eta(y) + \frac{\eta}{\gamma}\int_1^y \Omega(\psi,\gamma, \beta)\left[\frac{\textsc{P}_\eta(y)\textsc{Q}_{\eta}(\psi)-\textsc{Q}_\eta(y)\textsc{P}_{\eta}(\psi)}{\textsc{P}_{\tilde{\eta}}(\psi)\textsc{Q}_\eta(\psi) - \textsc{P}_\eta(y) \textsc{Q}_{\tilde{\eta}}(\psi)}\right]\mathrm{d}\psi~,
\end{equation}
where $\tilde{\eta}=1/2\left(1+\sqrt{1-4\gamma}\right)$~. Therefore, we may pursuit a wide family of solutions to Eq.~\eqref{EMGD2almostDE2} depending on the aforementioned parameters, which are crucial to estimate the final shape of the $\uprho_2(y)$ in Eq.~\eqref{eq:rho2EMGD2almost}. The constants of integration $\textsc{B}_1$ and $\textsc{B}_2$ depend on the computation of the integral carrying the $\Omega$-function.

Hereon we opt to work with two main scenarios. The first one consists to regard only the $1^{\rm st}$-order at $\varepsilon$, considering $\varepsilon^2\uprho_2(y)$ insignificant, compared to $\varepsilon\uprho_1(y)$. In fact, it is also consistent with the MGD and EMGD$_1$ scenarios, where $\uprho_2=0$. The second one goes to the $2^{\rm nd}$-order with some kind of simplifications to the $\Omega(y,\gamma,\beta)$ through free choice of values for the $\gamma$ and $\upbeta$ parameters to fit consistent solutions.

 \paragraph{First scenario: cutting off  $\varepsilon^2\uprho_2$.}

 In this case, only the $\varepsilon$-order for $\uprho$-function is imperative, leading us to deal with a simplified solution.
 
Next, it is important to expand the integrand in the Eq.~\eqref{area-minEMGD2almost}, which yields
\begin{equation}\label{eq:LIntegrandk=2almost}
\mathcal{L}= 2\pi\mathring{r}^2 + \left[4\pi\mathring{r}^2\frac{\textsc{P}_\eta(y)}{\textsc{P}_\eta(y_0)}\right]\varepsilon + \frac{\pi\mathring{r}^2}{\textsc{P}^2_\eta(y_0)}\left[\textsc{P}_\eta^2(y)-\frac{3}{\gamma}\left(y^2-1\right)\dot{\textsc{P}}_\eta^2(y)\right]\varepsilon^2 + \cdots
\end{equation}
and calculate the perturbative entropy function $\mathcal{S}=\mathcal{S}_0+\mathcal{S}_1+\mathcal{S}_2+\cdots$. It is crucial to keep in mind that we stand for up to second order.

Now, using only $\uprho_1(y)$, we determine the contributions to the entropy, order by order, up to the second one. Thus, the $0^{\rm th}$, $1^{\rm st}$ and $2^{\rm nd}$-orders are, respectively,
\begin{align}
    \mathcal{S}_0^{\scalebox{.5}{EMGD$_2$}}&=\frac{\pi\mathring{r}^2}{2}\left(1-y_0\right),\label{eq:S0EMGD2almost}\\
    \mathcal{S}_1^{\scalebox{.5}{EMGD$_2$}}&= \frac{\pi\uprho_0\mathring{r}}{\textsc{P}_\eta(y_0)}\int_{y_0}^1 \textsc{P}_\eta(y)\mathrm{d}y~,\label{eq:S1EMGD2almost}\\
    \mathcal{S}_2^{\scalebox{.5}{EMGD$_2$}}&=\frac{\pi\uprho_0^2}{2\textsc{P}_\eta^2(y_0)} \int_{y_0}^1\left[\textsc{P}_\eta^2(y)-\frac{3}{\gamma}\left(y^2-1\right)\dot{\textsc{P}}_\eta^2(y)\right]\mathrm{d}y~.\label{eq:S2EMGD2almost}
\end{align}

\begin{figure}[htb!]
    \centering
    \includegraphics[scale=.5]{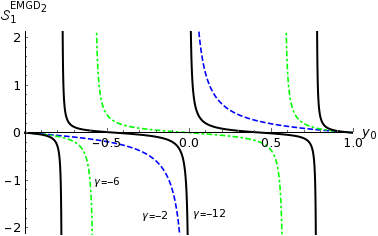}\quad\quad
    \includegraphics[scale=.5]{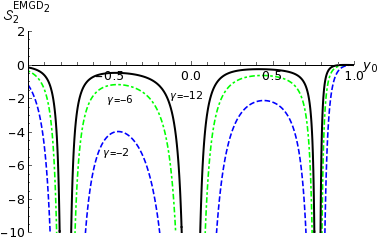}
    \caption{Behavior of the $1^{\rm st}$-order (on the left) and $2^{\rm nd}$-order (on the right) corrections of the HEE. For the former, the thick lines stand for $\gamma=-12$, with asymptotes at $y_0=0$, $y_0\approx -0.7746$ and $y_0\approx 0.0.7746$; the dot-dashed lines stand for $\gamma=-6$, with asymptotes at $y_0\approx-0.5774$; and $y_0\approxeq 0.5774$. The dashed lines stand for $\gamma=-2$ with a single asymptote at $y_0=0$. For the latter, all curves share the same asymptotes at $y_0=0$, $y_0\approx -0.7746$ and $y_0\approx 0.7746$.}
    \label{fig:S1and2EMGD2almost}
\end{figure}

A first novelty concerns about a non-vanishing $1^{\rm st}$-order correction for the HEE, which did not happened either in the MGD or in the EMGD$_1$ cases. The computation of a numerical value depends on the parameters $\gamma$ and $y_0$. Then we must plot  Eqs.~\eqref{eq:S1EMGD2almost} and \eqref{eq:S2EMGD2almost} considering some values for those parameters. Fig.~\ref{fig:S1and2EMGD2almost} shows three values for $\gamma$ -- the parameter gathering the $c$-coefficients with information about the ADM mass and the tidal charge as well. Meanwhile,  the range $-1<y_0<1$ is imposed, regarding the lower limit of integration that determines the size of the boundary.

\noindent On the one hand, there is a change of sign of the HEE $1^{\rm st}$-order correction between the asymptotes, for each value of $\gamma$. It indicates a substantial contribution from the EMGD parameters. On the other hand, we see only negative corrections at $2^{\rm nd}$-order correction.

It is worth to emphasize that chosen values for $\gamma$ generate the simplest polynomials as a manner to investigate a particular behavior of such order of correction. In a more realistic scenario, we will need precisely the physical values for both the ADM mass and the tidal charge, to fully  understand the contribution at this order. 

Additionally, in fact, for a set of values implying that $\Omega(\psi,\gamma,\beta)\to 0$ in Eq.~\eqref{eq:rho2EMGD2almost}, aggregating the boundary condition $\uprho_2(y_0)=0$ yields $\uprho_2(y)=0$. Under such circumstances, we obtain the same result found in this scenario.

\paragraph{Second scenario: samples for the $\Omega(y,\gamma,\beta)$ function.}

At this point, first, we choose two pair of values for $\gamma$ and $\beta$ to determine $\Omega(y,\gamma,\beta)$,  permitting us to determine the HEE corrections. Second and last, we attribute a value for $\gamma$ to find the corresponding numerical value for the EMGD$_2$ parameter dealing with a unit value for the mass parameter $M$.

As a first example, we take  $\gamma=-2$ and $\beta=0$. Hence the boundary condition $\uprho_1(y_0)=\mathring{r}$, Eq.~\eqref{EMGD2almostDE1},  provides $
    \uprho_1(y)=\frac{\mathring{r}y}{y_0}.$
These values also permit us to write
$
    \Omega(y,-2,0)=\frac{2\left(2-3y^2\right)\mathring{r}}{y_0^2}. $ Replacing it  into Eq.~\eqref{EMGD2almostDE2} yields
\begin{equation}
    \uprho_2(y)=~\frac{\mathring{r}}{2y_0^3}\left(y-y_0\right)\left(3yy_0-1\right).
\end{equation}
With the $\uprho$-functions, the expansion of the integrand in Eq.~\eqref{area-minEMGD2almost} can be found, resulting
\begin{equation}\label{LexpansionEMGD2almost2ndSc}
    \mathcal{L}=2\pi\mathring{r}^2+\left(\frac{4\pi\mathring{r}^2 y}{y_0}\right)\varepsilon+\frac{\pi\mathring{r}^2}{y_0^3}\left[2y_0+8y_0y^2-2y\left(1+3y_0^2\right)+3\left(y^2-1\right)y_0\right]\varepsilon^2+\cdots~.
\end{equation}

\begin{figure}[htb!]
    \centering
    \includegraphics[scale=.5]{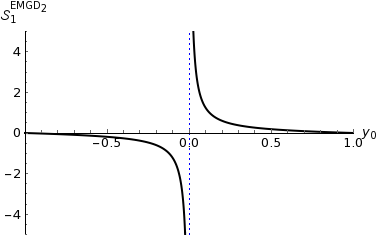}\qquad\quad
    \includegraphics[scale=.5]{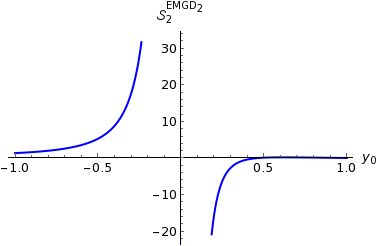}
    \caption{Profile of the $1^{\rm st}$-order (on the left) and $2^{\rm nd}$-order (on the right) corrections of the HEE for $\gamma=-2$ and $\beta=0$. In both plots, the asymptote is localized at $y_0=0$. The asymptotes are localized at $y_0=0$ for the both corrections.}
    \label{fig:S1andS2EMGD2almostEx1}
\end{figure}

The next step comprise to calculate the HEE corrections up to $2^{\rm nd}$-order, as implemented, employing  Eq.~\eqref{LexpansionEMGD2almost2ndSc}. Also, it is necessary to remember that $\varepsilon=\uprho_0/\mathring{r}$. Therefore, it implies that
\begin{align}
    \mathcal{S}_0^{\scalebox{.5}{EMGD$_2$}}&=\frac{\pi\mathring{r}}{2}\left(1-y_0\right)~,\\
    \mathcal{S}_1^{\scalebox{.5}{EMGD$_2$}}&=\frac{\pi\mathring{r}\uprho_0}{2y_0}\left(1-y_0^2\right)~,\\
    \mathcal{S}_2^{\scalebox{.5}{EMGD$_2$}}&=\frac{\pi\uprho_0^2}{4y_0^3}\left(-1+\frac{8y_0}{3}+5y_0^2-\frac{20y_0^4}{3}\right)~.
\end{align}
Fig.~\ref{fig:S1andS2EMGD2almostEx1} illustrates the last two outcomes above.
Once again, one can notice the appearance of the $1^{\rm st}$-order correction, which is not present in cases like the MGD or the EMGD$_1$. In addition, there is a sign change of such correction as well as can be observed in the case where $\uprho_2$ is insignificant.

As a second example, let us take $\gamma=-6$ and $\beta=8$. Similarly proceeding as in the previous example,  the $\uprho$-functions can be derived, as
\begin{align}
    \uprho_1(y)&=\frac{\mathring{r}\left(3y^2-1\right)}{\left(3y_0^2-1\right)}~.
\end{align}
Hence, one obtains $
    \Omega(y,-6,8)={42\mathring{r}\left(1-3y^2\right)^2}/{\left(1-3y_0^2\right)^2}$, yielding 
\begin{equation}
    \uprho_2(y)=~-\frac{9\mathring{r}\left(y^2-y_0^2\right)}{5\left(3y_0^2-1\right)^3}\left[13-15y_0^2+15y^2\left(3y_0^2-1\right)\right].
\end{equation}
One more time, with these $\uprho$-functions, we expand the integrand in Eq.~\eqref{area-minEMGD2almost}, which leave us with
\begin{align}\label{LexpansionEMGD2almost2ndEx2}
    \mathcal{L}&=2\pi\mathring{r}^2+\left[\frac{4\pi\mathring{r}^2\left(3y^2-1\right)}{3y_0^2-1}\right]\varepsilon+\frac{2\pi\mathring{r}^2}{5\left(3y_0^2-1\right)^3}\left[-5+249y_0^2-270y_0^4-225y^4\left(3y_0^2-1\right)\right.\nonumber\\
    &\left.\quad+6y^2\left(-34-15y_0^2+135y_0^4\right)+279y^2\left(y^2-1\right)\left(3y_0^2-1\right)\right]\varepsilon^2+\cdots~.
\end{align}

\begin{figure}[htb!]
    \centering
    \includegraphics[scale=.5]{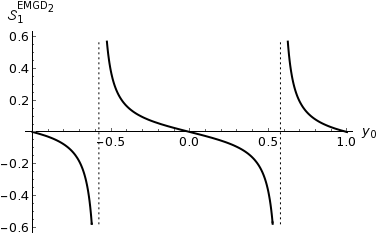}\quad\quad
    \includegraphics[scale=.5]{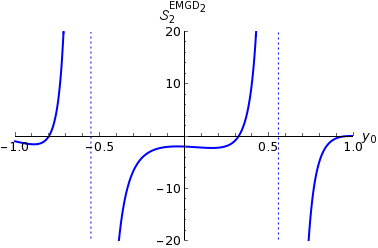}
    \caption{Profile of the $1^{\rm st}$-order (on the left) and $2^{\rm nd}$-order (on the right) corrections of the HEE for $\gamma=-6$ and $\beta=8$. For both orders, the asymptotes are situated at $y_0=\mp0.5773$.}
    \label{fig:S1andS2EMGD2almostEx2}
\end{figure}

Finally, after employing Eq.~\eqref{area-minEMGD2almost}, the calculations of the HEE corrections, order by order up to the second one, read
\begin{align}
    \mathcal{S}_0^{\scalebox{.5}{EMGD$_2$}}&=\frac{\pi\mathring{r}}{2}\left(1-y_0\right)~,\\
    \mathcal{S}_1^{\scalebox{.5}{EMGD$_2$}}&=\pi\mathring{r}\uprho_0\left(\frac{y_0 - y_0^3}{3y_0^2-1}\right)~,\\
    \mathcal{S}_2^{\scalebox{.5}{EMGD$_2$}}&=\frac{2\pi\uprho_0^2}{5}\left[\frac{8 + 5y_0 - 24y_0^2 - 271y_0^3 + 579y_0^5 - 297y_0^7}{\left(-1 + 3 y_0^2\right)^3}\right]~.
\end{align}
Fig.~\ref{fig:S1andS2EMGD2almostEx2} brings the profile of the last two entropy functions.

 The appearance of the $1^{\rm st}$-order correction happens again, with the sign-changing noticed before in the first example.

As a third example, we adopt the mass parameter $M=1$, which leaves us with $\mathbb{M}_2=3M=3$, $\mathbb{Q}_2=12M^2=12$ and $\mathring{r}=3.36$. Once  the numerical value of $\textsc{s}$ is not available to  determine entirely the $c$-coefficients in \eqref{k=2Coeffs}, then $\gamma$ is fixed with respect to a well known Legendre polynomial. With the choice $\gamma=-20$, hence, we figure out that $\textsc{s}\approx -17.9841$ and, consequently, $\beta\approx 3.3942$. The underlying computation is usual with truncations made on the numerical values for all parameters, up to four decimal places.

Therefore, performing strictly as in the previous two examples, we obtain the first $\uprho$-function as a solution of Eq.~\eqref{EMGD2almostDE1}, that is,
\begin{equation}
    \uprho_1(y)=3.36\left(\frac{3-30y^2+35y^4}{3-30y_0^2+35y_0^4}\right)~.
\end{equation}
With the numerical values, then
\begin{equation}
    \Omega(y,-20,3.3942)=\frac{1.1820-0.9015y^2+1.9291y^4-45.8903y^6+37.0863y^8}{\left(0.0857-0.8571y_0^2+y_0^4\right)^2}\nonumber
\end{equation}
permits to determine the second $\uprho$-function solving the Eq.~\eqref{EMGD2almostDE2}, \emph{i.e.},
\begin{equation}
    \uprho_2(y)=-\frac{0.7131\left(y^2-y_0^2\right)}{\mathrm{p}_6^3}\sum_a \mathrm{p}_ay^a~,
\end{equation}
where $a\in\left\{0,2,4,6\right\}$ and
\begin{align}
\mathrm{p}_0&=0.0857 y_0^6 - 0.0325 y_0^4 + 0.0439 y_0^2 - 0.0659,\nonumber\\
\mathrm{p}_2&=-0.8571 y_0^6 + 0.4108 y_0^4 + 0.2975 y_0^2 + 0.0439,\nonumber\\
\mathrm{p}_4&=y_0^6 - 1.2364 y_0^4 + 0.4108 y_0^2 - 0.0325,\nonumber\\
\mathrm{p}_6&=y_0^4-0.8571 y_0^2 + 0.0857.\nonumber
\end{align}

Now, we proceed with the expansion of the integrand in Eq.~\eqref{area-minEMGD2almost} to help us to determine the HEE corrections, which yields 
\begin{equation}
\mathcal{L} =\mathcal{L}_0 + \left[4.7286\times 10^1\left(\frac{3-30y^2+35y^4}{3-30y_0^2+35y_0^4}\right)\right]\varepsilon + \left[\left(\frac{1.12}{\mathrm{D}_0^2}\right)^2\sum_{a=0}^{14}\sum_{b=0}^4\mathrm{N}_{ab}\left(y_0^2\right)^a\left(y^2\right)^b\right]\varepsilon^2 + \cdots~,\label{LexpandedEMGD2almostSampl3}
\end{equation}
with $\mathcal{L}_0=7.0934\times 10$~. The numerical coefficients $\mathrm{N}_{ab}$ are displayed  in~\ref{appendixA}. Meanwhile,
\begin{equation}
    \mathrm{D}_0=\sum_{i=0}^4 \mathrm{D}_{0i}(y_0^2)^i~,\nonumber
\end{equation}
with $\mathrm{D}_{00}=7.3457\times 10^{-3}, \mathrm{D}_{01}=1.4692\times 10^{-1}, \mathrm{D}_{02}=9.0607\times 10^{-1}$~, $\mathrm{D}_{03}=-1.7142$ and $\mathrm{D}_{04}=1$.

With the integrand in hands, we can compute the HEE corrections, order by order, up to second one, as follows
\begin{subequations}
\begin{align}
    \mathcal{S}_0^{\scalebox{.5}{EMGD$_2$}}&=17.7337\left(1-y_0\right)~,\label{S0EMGD2almostSample3}\\
    \mathcal{S}_1^{\scalebox{.5}{EMGD$_2$}}&=~\left(\frac{- 0.9047y_0 + 3.0159y_0^3 - 2.1111y_0^5}{0.0857 - 0.8571y_0^2 + y_0^4}\right)\uprho_0,\label{S1EMGD2almostSample3}\\
    \mathcal{S}_2^{\scalebox{.5}{EMGD$_2$}}&=\frac{\uprho_0^2}{\mathrm{D}_0^4}\sum_{i=0}^{33}\mathrm{K}_i y_0^i~.\label{S2EMGD2almostSample3}
\end{align}
\end{subequations}
where the numerical coefficients $\mathrm{K}_i$ are listed in \ref{appendixB}. Fig.~\ref{fig:S1andS2EMGD2almostEx3} shows the shape of the last two entropy functions above.
\begin{figure}[htb!]
    \centering
    \includegraphics[scale=.5]{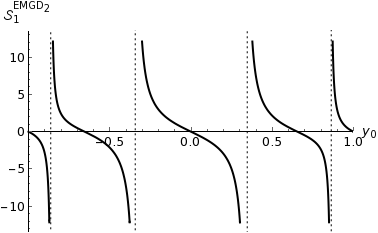}\qquad\qquad
    \includegraphics[scale=.5]{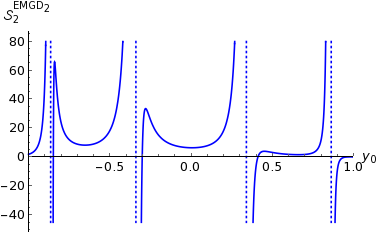}
    \caption{Profile of the $1^{\rm st}$-order (on the left) and $2^{\rm nd}$-order (on the right) corrections of the HEE for $\gamma=-20$, $\textsc{s}\approx -17.9841$ and $\beta=3.3942$. Both the plots display their asymptotes at $y_0=\mp0.8611$ and $\mp y_0=-0.3399$.}
    \label{fig:S1andS2EMGD2almostEx3}
\end{figure}
The profile of the $1^{\rm st}$-order correction has a sign-changing noticed before in both previous examples. Besides, there is an alternate behavior looking at the two last corrections. Now, there is a sign-changing with an attenuation  in the increment of the values for both corrections.

\section{Conclusions}\label{sect:Finals}

\quad\;About the MGD case, we calculated the HEE of the MGD solution to investigate the influence of high energy effects caused by the MGD parameter $\ell$, encoded in the $\xi$ parameter, from the AdS/CFT membrane paradigm. There are two perspectives, namely, the almost on the horizon and far from the horizon regimes. Far from the horizon, the HEE $0^{\rm th}$-order is not affected by $\xi$, which is a good feature of the deformation, as Eq.~\eqref{S0MGDbeyond} exactly matches the HEE for Schwarzschild spacetimes, as pointed out in Ref. \cite{Sun:2016dch}. The novelty clearly appears when one reaches the HEE $1^{\rm st}$-order correction, since the $\xi$ parameter is present in Eq.~\eqref{S1MGDbeyond} as well as in the ratio casted by Eq.~\eqref{phi1MGD}. The fact that $\xi<0$, due to the same sign of $\ell$, contributes to an increment of the correction term, however without any modification of its sign, which is made explicit by Eq.~\eqref{phi1MGD}. Once more, the MGD parameter carrying on brane effects is featured in the HEE $2^{\rm nd}$-order correction, as revealed by Eqs.~\eqref{S2MGDbeyond}.
Computations in this direction shed new light about holography in asymptotically flat spaces.


Comparatively with the HEE for Schwarzschild black hole, one notices the exponential rise of such order of correction, when the brane tension  is lowered, as illustrated by Fig.~\ref{fig:phi2MGD}. Therefore, lower brane tension values have profound influence in the increment of this order of correction, as one can see in Fig.~\ref{fig:S2MGD}. Another feature is the agreement with the first law for the  HEE,  evinced by  Eqs.~\eqref{S1MGDbeyond} and \eqref{S2MGDbeyond}. Fig.~\ref{fig:S2MGD_M_and_xi} shows that the more the MGD black hole mass increases, the higher the magnitude of the $2^{\rm nd}$-order correction is, concomitantly to the rise of the size of the subsystem, which is characterized by $y_0$. Fig.~\ref{fig:S2Schwarz} permits us to obtain a better comparison of this feature, while one looks at the HEE $2^{\rm nd}$-order correction for a Schwarzschild black hole. As expected, when $\xi=0$, the HEE corrections for Schwarzschild spacetimes are recovered at $1^{\rm st}$- and $2^{\rm nd}$-order, accordingly. Even when one considers the boundary far away from the event horizon of the MGD black hole, it is observed substantial differences when confronted to the HEE of a typical Schwarzschild black hole.
 
 

Regarding the entangling surface almost on the horizon, the MGD parameter, $\ell$, that encodes the finite brane tension, demonstrated its strength to modify the HEE $2^{\rm nd}$-order correction, as one can notice in  Fig.~\ref{fig:S2MGDalmost}. The MGD influence is codified by the parameter $\alpha$, which is correspondent, in a brief mode, to $\ell$. The $0^{\rm th}$-order and $1^{\rm st}$-order are not susceptible to such parameter and both of them match to those ones established in Ref. \cite{Sun:2016dch}. On the other hand, the low brane tension weighs significantly to lift the maximum value of the HEE $2^{\rm nd}$-order correction,  according to Fig.~\ref{fig:S2MGDalmost}. Here, so close to the event horizon of a MGD black hole, the correction at $2^{\rm nd}$-order is more sensitive to the MGD parameter.


Extending the analysis from the Schwarzschild black holes and illustrated by Fig.~\ref{fig:S2MGDalmost}, an important novelty consists of the maximum value arising in the MGD HEE $2^{\rm nd}$-order correction. This is associated with lowering the brane tension and, concomitantly, requires a large size of the dual quantum subsystem.

In the EMGD$_1$ case, a similar scenario to the Reissner-Nordstr\"om spacetime occurs. Far from the horizon, similarly to  the HEE for the Schwarzschild spacetime, a subtle numerical shift of the HEE $1^{\rm st}$-order correction is verified with  $\Phi_1^{\scalebox{.5}{EMGD$_1$}}=2$. It happens due to the presence of the ADM mass in Eq.~\eqref{eq:S1EMGD}. Meanwhile, the $0^{\rm th}$-order is not altered. It is worth to emphasize that the correspondence between the tidal charge, $\mathbb{Q}_1$, and the ADM mass, $\mathbb{M}_1$, with the Schwarzschild mass $M$, is mostly necessary to analyze the relative behavior of the HEE corrections for the EMGD$_1$ spacetime. The ratio \eqref{eq:phi2EMGD1} shows the peculiarity of such correspondence, which is sustained by Fig.~\ref{fig:phi2EMGD1}.  The influence of the black hole mass is notorious with the large size of the entangling surface characterized by $y_0$, as shown by Fig.~\ref{fig:S2EMGD1_M_and_y0}. The increments in the HEE $2^{\rm nd}$-order correction are accentuated accordingly with the mass increment and the enlargement of the minimal area. Figs.~\ref{fig:S2Schwarz_EMGD1_great_and_small_y} and \ref{fig:S2EMGD1_M_and_y0} make us to comprehend that the greater the mass, the greater the deviation of the HEE $2^{\rm nd}$-order correction for the EMGD$_1$ is, related to that one for the Schwarzschild spacetime.



 Considering the entangling surface almost on the horizon for the $k=1$ scenario, we have only an extremal black hole with the degenerate horizon $\mathring{r}=\mathbb{M}_1$. The HEE for this case is very close to the HEE for the Reissner-Nordstr\"om spacetime. The crucial distinction relies on the numerical value of the full entropy displayed by Eq.~\eqref{eq:SEMGD1almost} in consequence of the weakening of the gravitational field carried by the position of the event horizon in an EMGD$_1$ spacetime, that is, $\mathring{r}=\mathbb{M}_1=2M$. The relationship between those  entropies displays a limit equal to $4$ and it is sustained by Eq.~\eqref{eq:ratioSEMGD1toRN} and exhibited by Fig.~\ref{fig:ratioSEMGD1almost}, where it is possible to notice a fast-growing ratio as the mass of the black hole increases.

The EMGD$_2$ case brings on the possibility to settle additional HEE corrections to a certain class of black holes beyond Reissner-Nordstr\"om spacetimes. Far from the horizon, the HEE $0^{\rm th}$-order is not affected, behaving  like a constant, as  the HEE for all cases are confronted. As occurred in the MGD case, the HEE $1^{\rm st}$-order correction displays already the specific quantifier related to the brane tension, \emph{i.e.}, the parameter $\textsc{s}$, as shown by Eq.~\eqref{S1EMGD2beyond}. Besides, the HEE $2^{\rm nd}$-order correction is richer, despite its structural similarity when faced up to the same order in either the MGD or the EMGD$_1$ cases. The mass terms are  preserved, which is a welcome feature to hold the first law of HEE. The new establishment has tuned with the quadratic term in $\textsc{s}$ and the mixed one with $\mathbb{M}_2$ and $\textsc{s}$, as supported by Eq.~\eqref{S2EMGD2beyond}.

The $\Phi$-ratios were also computed, scaling with Schwarzschild mass $M$. In Fig.~\ref{fig:S2EMGD2_M_zeta-01and-100}, we observe two simple scenarios fixing the brane tension parameter. It  unveils the fast-growing of the HEE $2^{\rm nd}$-order correction according to the mass parameter and the size of the minimal area. Fig.~\ref{fig:phi2EMGD2} exposes how the brane tension affects, relatively, the HEE $2^{\rm nd}$-order correction, where it is clear that lower tension branes have exponential gains, consonantly with the size of the minimal area, that is, the range of the dual subsystem that entanglement entropy stands for. In addition, looking at Fig.~\ref{fig:S2EMGD2_S2Schw_M_and_zeta-01_y0_closeto0and1} and Fig.~\ref{fig:S2EMGD2_S2Schw_M_and_zeta-100_y0_closeto0and1}, one can notice the significant deviation between the HEE for a Schwarzschild black hole and the HEE for the EMGD$_2$ spacetime.  For completeness, Fig.~\ref{fig:phi2EMGD2_zeta_y0} shows how the $2^{\rm nd}$-order ratio behaves under the simultaneous variation of the brane tension and the size of the dual subsystem.

With the entangling surface almost on the horizon, we employ the expansion of an auxiliary function characterizing the proximity to the horizon which head us to general analytical solutions depending on parameters related to the ADM mass, tidal charge, and brane tension. Therefore, we expend efforts to analyze some possible scenarios towards the profile of HEE corrections in this present case.
Firstly, based on a meaningless $\uprho_2$, we determine HEE corrections very similar to the previous cases, \emph{i.e.}, MGD and EMGD$_1$, as shown by Eqs.~\eqref{eq:S0EMGD2almost} and \eqref{eq:S2EMGD2almost}. The dependence of the starting point at horizon $\uprho_0$ is sustained at $1^{\rm st}$-order and $2^{\rm nd}$-order corrections. In addition, this approximation requires using values to $\gamma$ and the plots in Fig.~\ref{fig:S1and2EMGD2almost}, even dealing with simple Legendre polynomials, displaying the sign-changing demeanor of the two orders of corrections for the HEE. Of course, if we limit ourselves to a certain region into the boundaries, which means to limit the size of the dual subsystem, we get away from the asymptotic regions. Besides, among the asymptotes we observe the similar behavior of both HEE orders of corrections. Secondly, we scrutinize three examples, each one of them demonstrates sign-changing behavior of the HEE $1^{\rm st}$- and $2^{\rm nd}$-order corrections. The $0^{\rm th}$-order, as usual, remains immutable. According to the rank of the Legendre polynomials corresponding to the choices for $\beta$ and $\gamma$, we handled with one to two asymptotes marking the regions where the change of sign of that order of correction occurs,  as one can realize in Figs.~\ref{fig:S1andS2EMGD2almostEx1} and \ref{fig:S1andS2EMGD2almostEx2}. The last example was built attributing a mass reference and, subsequently, fixing the brane tension parameter, $\textsc{s}$, with determined value for $\gamma$, which is purposely attached to the order of a rank-4 Legendre polynomial. Its functionality as a toy model reveals the same sign-changing aspect of the orders of corrections for the respective HEE. The new aspect noticed here was in virtue to the local maxima and minima presented at HEE $2^{\rm nd}$-order correction as showed by Fig.~\ref{fig:S1andS2EMGD2almostEx3}. Such presence of extremal points reveals a real constraint to the corrections for the HEE. Finally, specific values for the physical parameters bring to us the most realistic results for the HEE in EMGD$_2$ spacetimes. Without lose of clarity the constructions of the toy models aforementioned was essential to the simplest landscapes.
 

\paragraph*{Acknowledgments:}\; \;

   RdR~is grateful to FAPESP (Grant No.  2017/18897-8) and to the National Council for Scientific and Technological Development  -- CNPq (Grants No. 406134/2018-9 and No. 303293/2015-2), for partial financial support. AAT thanks to PNPD -- CAPES -- UFABC (Proc. No. 88887.338076/2019-00). 

\appendix

\section{The numerical coefficients -- Part I}\label{appendixA}

\quad\;To simplify, the $\mathrm{N}_{ab}$ parameters in Eq.~\eqref{LexpandedEMGD2almostSampl3}, $a\in\left\{0,1,2,\dots,14\right\}$ and $b\in\left\{0,1,\dots,4\right\}$, are displayed as a matrix form  below, wherein $a$ stands for rows and $b$ for columns.
\begin{equation}\small
\begin{bmatrix*}[r]
1.8294\times 10^{-8} & -6.4935\times 10^{-7} & 7.9329\times 10^{-6} & -1.4964\times 10^{-5} & 7.4078\times 10^{-6} \\
- 1.9118\times 10^{-6} & 4.7106\times 10^{-5} & -4.8551\times 10^{-4} & 8.9794\times 10^{-4} & -4.4451\times 10^{-4} \\
6.9972\times 10^{-5} & -1.4322\times 10^{-3} & 1.2938\times 10^{-2} & -2.3498\times 10^{-2} & 1.1632\times 10^{-2} \\
-1.3191\times 10^{-3} & 2.4156\times 10^{-2} & -1.9684\times 10^{-1} & 3.5175\times 10^{-1} & -1.7413\times 10^{-1} \\
1.4700\times 10^{-2} & -2.5062\times 10^{-1} & 1.8861 & -3.3237 & 1.6454 \\
-1.0277\times 10^{-1} & 1.6725 & -1.1869\times 10^{1} & 2.0683\times 10^{1} & -1.0239\times 10^{1} \\
4.6293\times 10^{-1} & -7.3164 & 4.9859\times 10^{1} & -8.6183\times 10^{1} & 4.2664\times 10^{1} \\
-1.3661 & 2.1186\times 10^{1} & -1.4043\times 10^{2} & 2.4133\times 10^{2} & -1.1947\times 10^{2} \\
2.7082 & -4.1192\times 10^{1} & 2.6502\times 10^{2} & -4.525\times 10^{2} & 2.2401\times 10^{2} \\
-3.7624 & 5.5047\times 10^{1} & -3.3213\times 10^{2} & 5.5875\times 10^{2} & -2.7661\times 10^{2} \\
3.9379 & -5.296\times 10^{1} & 2.7061\times 10^{2} & -4.3552\times 10^{2} & 2.156\times 10^{2} \\
-3.3659 & 3.9715\times 10^{1} & -1.3944\times 10^{2} & 1.9419\times 10^{2} & -9.6133\times 10^{1} \\
2.3127 & -2.4305\times 10^{1} & 4.6461\times 10^{1} & -3.776\times 10^{1} & 1.8693\times 10^{1} \\
-1.0665 & 1.0665\times 10^{1} & -1.2443\times 10^{1} & 0 & 0\\
2.2863\times 10^{-1} & -2.2863 & 2.6674 & 0 & 0\nonumber
\end{bmatrix*}
\end{equation}

\section{The numerical coefficients -- Part II}\label{appendixB}

With $i\in\{0,1,2,\dots,32,33\}$, the numerical coefficients $\mathrm{K}_i$ for the Eq.~\eqref{S2EMGD2almostSample3} are
$$
\begin{smallmatrix*}[l]
    \mathrm{K}_0=1.8435\times 10^{-8} & \mathrm{K}_1=-4.5735\times 10^{-9} & \mathrm{K}_2=-1.1061\times 10^{-6} & \mathrm{K}_3=5.3205\times 10^{-7} & \mathrm{K}_4=2.8944\times 10^{-5} \\
    \mathrm{K}_5=-2.1815\times 10^{-5} & \mathrm{K}_6=-4.3326\times 10^{-4} & \mathrm{K}_7=-4.7394\times 10^{-4} & \mathrm{K}_8=4.0938\times 10^{-3} & \mathrm{K}_9=-6.3671\times 10^{-3} \\
    \mathrm{K}_{10}=-2.5474\times 10^{-2} & \mathrm{K}_{11}=5.7270\times 10^{-2} & \mathrm{K}_{12}=1.0614\times 10^{-1} & \mathrm{K}_{13}=-3.6230\times 10^{-1} & \mathrm{K}_{14}=-2.9721\times 10^{-1} \\
    \mathrm{K}_{15}=1.6682 & \mathrm{K}_{16}= & \mathrm{K}_{17}=-5.7199 & \mathrm{K}_{18}=-6.8807\times 10^{-1} & \mathrm{K}_{19}=1.4757\times 10^{1} \\
    \mathrm{K}_{20}=5.3631\times 10^{-1} & \mathrm{K}_{21}=-2.8627\times 10^{1} & \mathrm{K}_{22}=-2.3912\times 10^{-1} & \mathrm{K}_{23}=4.1341\times 10^{1} & \mathrm{K}_{24}=4.6496\times 10^{-2} \\ 
    \mathrm{K}_{25}=-4.3596\times 10^{1} & \mathrm{K}_{26}=1.1102\times 10^{-16} & \mathrm{K}_{27}=3.2502\times 10^{1} & \mathrm{K}_{28}=7.6328\times 10^{-17} & \mathrm{K}_{29}=-1.6193\times 10^{1} \\
    \mathrm{K}_{30}=0.0000 & \mathrm{K}_{31}=4.8316 & \mathrm{K}_{32}=0.0000 & \mathrm{K}_{33}=-6.5262\times 10^{-1} &  \\
\end{smallmatrix*}\nonumber
$$

\bibliographystyle{spphys}
\bibliography{1bib_HEE}

\end{document}